\documentclass[fleqn,5p]{elsarticle}

\journal{Sustainable Energy, Grids and Networks}

\usepackage{fix-cm}
\usepackage{multicol}
\usepackage{amsmath}
\usepackage{amssymb}
\usepackage{mathtools}
\usepackage{upgreek}
\usepackage{cuted}
\usepackage{graphicx}
\usepackage{flafter}
\usepackage{placeins}
\usepackage{IEEEtrantools}

\usepackage{textcomp}

\usepackage{hyperref}
\usepackage{url}
\usepackage{color}
\usepackage{tabularx}
\usepackage{booktabs}
\usepackage{multirow}
\usepackage{pgfplots}
\usepackage{xcolor}
\usepackage[table]{xcolor}
\usepackage{caption}

\allowdisplaybreaks
\pgfplotsset{compat=1.18}

\hypersetup{urlcolor=cyan}

\begin{document}
%\linenumbers

\setlength{\tabcolsep}{6pt}
\setlength{\mathindent}{0pt}

\begin{frontmatter}

\title{Imperfect Competition in Markets for Short-Circuit Current Services}

\author[UPM]{P. Wang}
\ead{peng.wang@alumnos.upm.es}
\author[UPM]{L. Badesa}
\ead{luis.badesa@upm.es}
\address[UPM]{School of Industrial Engineering and Design (ETSIDI), Technical University of Madrid (UPM), Spain}

\begin{abstract}
An important limitation of Inverter-Based Resources (IBR) is their reduced contribution to Short-Circuit Current (SCC), as compared to that of Synchronous Generators (SGs). With increasing penetration of IBR in most power systems, the reducing SCC poses challenges to a secure system operation, as line protections may not trip when required. In order to address this issue, the SCC ancillary service could be procured via an economic mechanism, aiming at securing adequate SCC on all buses. However, the suitability of markets for SCC services is not well understood, given that these could be prone to market power issues: since the SCC contributions from various SGs to a certain bus are determined by the electrical topology of the grid, this is a highly local service. It is necessary to understand if SGs at advantageous electrical locations could exert market power and, if so, how it could be mitigated. In order to fill this gap, this paper, for the first time, adopts an SCC-constrained bilevel model to investigate strategic behaviors of SGs. To address the non-convexity due to unit commitment variables, the model is restructured through a primal-dual formulation. Based on a modified IEEE 30-bus system, cases with strategic SGs placed at different buses are analyzed. These studies demonstrate that strategic agents exerting market power by manipulating service prices and extending operating periods could achieve up to triple revenues from SCC provision, which reduces market efficiency and would increase the financial burden on consumers. These findings highlight the need for careful market design, for which potential measures to mitigate these market power issues are also discussed.

\end{abstract}


\end{frontmatter}

\section*{Nomenclature}
\addcontentsline{toc}{section}{Nomenclature}

\subsection*{Abbreviations and Acronyms}
\begin{IEEEdescription}
    \item[IBR] Inverter-Based Resources
    \item[KKT] Karush–Kuhn–Tucker
    \item[LL] Lower-Level
   % \item[MPEC] ~ \parbox[t]{.9\linewidth} {Mathematical Program with Equilibrium Constraints}
    \item[SCC] Short-Circuit Current
    \item[SG] Synchronous Generator
    \item[UC] Unit Commitment
    \item[UL] Upper-Level

\end{IEEEdescription}

\subsection*{Indices and Sets}
\begin{IEEEdescription}
    \item[$b, \mathcal{B}$] \qquad Index, Set of buses
    \item[$c, \mathcal{C}$] \qquad Index, Set of IBR   
    \item[$g,\mathcal{G}$] \qquad Index, Set of generic SGs
    \item[$\hat{g},\hat{\mathcal{G}}$] \qquad Index, Set of strategic SGs
    \item[$\check{g}, \check{\mathcal{G}}$] \qquad Index, Set of non-strategic SGs   
    \item[$m, \mathcal{M}$]  \qquad Index, Set of commitment decision pairs
    \item[$t, T$]  \qquad Index, Set of time periods
    \item[$\Psi(g), \Phi(c)$]  \qquad Index of bus for SGs and IBR
    \item[$\omega, \Omega$]  \qquad Index, Set of system operating points 
\end{IEEEdescription}

%\vspace*{-6pt}
\subsection*{Constants and Parameters}
\begin{IEEEdescription}
    \item[$\upalpha_{c,t}$] Capacity factor of IBR at time period $t$
    \item[$\overline{\upbeta}^\textrm{m}_{\hat{g}}$] Upper bound of the strategic bidding multiplier for SGs in energy market 
    \item[$\overline{\upbeta}^\textrm{SCC}_{\hat{g},b}$] Upper bound of the strategic bidding multiplier for SGs in SCC service market of bus $b$
    \item[$\textrm{I}_{b_\textrm{lim}}$] Minimum requirement of SCC for bus $b$ (p.u.)
    \item[$\textrm{K}^\textrm{st}_g,\textrm{K}^\textrm{sh}_g$] \quad Start-up and shut-down costs of SGs (\texteuro/h) 
        \vspace{2pt}
    \item[$\textrm{k}_{bg}, \textrm{k}_{bc}, \textrm{k}_{bm}$] \qquad \quad  \parbox[t]{.85\linewidth} {Approximate coefficients of the SCC on bus $b$ (p.u.)}
    \item[$\textrm{O}_g^\textrm{nl}$]  No-load costs of SGs (\texteuro/h)
    \vspace{2pt}
    \item[$\textrm{O}_{g,b,t}^\textrm{SCC}$] Commitment price of SGs for providing SCC to bus $b$ at time period $t$ (\texteuro/h)
    \item[$\textrm{O}_g^\textrm{m}$]  Marginal power generation costs of SGs (\texteuro/MWh)  
    \vspace{2pt}
    \item[$\textrm{O}_c^\textrm{l}$] Energy bids of IBR (\texteuro/MWh) 
        \vspace{2pt} 
    \item[$\textrm{P}^\textrm{D}_t$] Total system demand at time period $t$ (MW) 
\item[$\textrm{P}_g^\textrm{min},\textrm{P}_g^\textrm{max}$] \qquad  \parbox[t]{.9\linewidth} { Minimum stable generation and rated power of SGs (MW)} 
        %\vspace{2pt}
\item[$\textrm{P}_g^\textrm{rd},\textrm{P}_g^\textrm{ru}$] \quad Ramp-down and ramp-up limits of SGs (MW/h) 
        \vspace{2pt}
\item[$\textrm{P}^{\textrm{max}}_{c}$] \quad Rated power of IBR (MW) 
    \vspace{2pt}
\item[W] \quad Penalty factor in the primal-dual formulation
\end{IEEEdescription}

\subsection*{Primal Variables}
\begin{IEEEdescription}
    \item[$\beta_{\hat{g},t}^\textrm{m}$]  Bidding multiplier of strategic SGs in energy market at time period $t$
    \item[$\beta_{\hat{g},b,t}^\textrm{SCC}$]  Bidding multiplier of strategic SGs in SCC service market of bus $b$ at time period $t$
    \item[$C_{g,t}^\textrm{st}, C_{g,t}^\textrm{sh}$] \quad~ \parbox[t]{.9\linewidth} {Start-up/shut-down costs incurred by SGs at time period $t$ (\texteuro/h)}
    \item[$P_{g,t}$]  Power output of SGs at time period $t$ (MW)
    \item[$P_{c,t}$] Power output of IBR at time period $t$ (MW)
    \item[$u_{g,t}$] Binary variable, commitment of SG at time period $t$
    %\item[$I_{F_\textrm{L}}$]  Approximate SCC contribution on bus $F$ (p.u.)
\end{IEEEdescription}

\subsection*{Dual Variables}
\begin{IEEEdescription}
    \item[$\lambda^\textrm{E}_{t}$] Energy clearing price at time period $t$ (\texteuro/MWh) 
    \item[$\lambda_{b,t}^{\textrm{SCC}}$]  SCC clearing price for bus $b$ at time period $t$ (\texteuro/p.u.)%(\texteuro/$\textrm{I}_{F_{\textrm{lim}}}$)
    \item[${\mu}^{\textrm{min}}_{g,t}, {\mu}^{\textrm{max}}_{g,t}$] \qquad Associated with the generation limits of SGs
    \item[$\pi_{g,t}^{\textrm{rd}}, \pi_{g,t}^{\textrm{ru}}$] \qquad Associated with the ramp constraints of SGs 
    \item[$\rho_{g,t}^{\textrm{st}}, \sigma_{g,t}^{\textrm{st}}, \rho_{g,t}^{\textrm{sh}}, \sigma_{g,t}^{\textrm{sh}}$] \qquad \qquad \quad \parbox[t]{.75\linewidth} { Associated with the constraints for start-up/shut-down costs of SGs }
    \item[$\zeta_{c,t}^{\textrm{min}},\zeta_{c,t}^{\textrm{max}}$] \qquad \parbox[t]{.9\linewidth} {  Associated with the constraints for generation limits of IBR }
    \item[$\psi_{g,t}^{\textrm{min}}, \psi_{g,t}^{\textrm{max}}$] \qquad \parbox[t]{.9\linewidth} { Associated with the constraints for relaxation of commitment status of SGs}
\end{IEEEdescription}

\section{Introduction} \label{Intro}
\subsection{Background and Motivation }
Electric power systems throughout the world are shifting to renewable energy dominated architectures for achieving the goal of net-zero carbon emissions in coming decades. This requires the massive installation of Inverter-Based Resources (IBR), mainly wind and solar generators. 

Since power electronic inverters have very limited overload capacity, they only contribute to about one-fifth of the Short-Circuit Current (SCC) provided by Synchronous Generators (SGs) \cite{tleis20197,jia2017review}. Consequently, the rapid integration of IBR significantly reduces system-wide SCC levels. Sufficient SCC is a main criterion for secure system operation \cite{dozein2018system,huuhtanen2024system}, as lack of this physical quantity can have important negative impacts, including causing protection devices to fail to respond to a short-circuit fault, especially on distant or lightly loaded feeders. Deficiency in SCC may also lead to voltage oscillations, potentially causing the phase-locked loop to lose synchronization. Such practical grid security concerns create a critical need to address SCC adequacy.

Various proposals have been made to overcome these issues in IBR-dominated systems, including inserting phase-change material into chips to improve the overload capability of grid inverters \cite{shao2020power}, and installing synchronous condensers to maintain the minimum SCC of critical buses \cite{jia2018synchronous}. Another feasible solution is considering SCC as an ancillary service, through dispatching SGs to offer sufficient SCC during certain periods, thus guaranteeing an adequate performance of protections during short-circuit faults \cite{chu2024pricing}. For power systems with both SGs and IBR, a methodology for evaluating SCC on specific buses within dispatch models was developed in \cite{chu2021short}. It computes the SCC contributions according to the Unit Commitment (UC) status and equivalent impedance between thermal units and buses concerned. It is important to note that online SGs provide little SCC to distant buses, and only contribute significantly to increasing SCC in their surrounding nodes. Therefore, once the SCC requirement of all buses is satisfied through dispatching certain critically-located SGs, the service provided by other ones will no longer be needed. This highly local nature of the SCC service may give certain generation companies some room to exert market power for increasing their own profits, particularly when they manage SGs located at advantageous electrical locations.

Against this backdrop, this paper aims to explore two challenges in SCC service procurement: First, to examine potential strategic behaviors of SGs in SCC service markets; Second, to quantify the extent of market power exploitation by advantageous-location SGs. Analyzing such market power issues in SCC services could provide insights into the following concerns:
\begin{enumerate}
    \item The markets for other stability services, such as frequency reserve and inertia, are already established in several real-world systems, such as Great Britain \cite{matamala2025strategic}. Thus, analyzing the potential market power risks in an SCC market is practically relevant today, as it could address a key need for SCC in modern power systems that can be met through market mechanisms.
    \item If such risks exist, how they could be mitigated or avoided.
\end{enumerate}

\subsection{Review of Relevant Studies}
Market power issues in the electricity industry have been extensively analyzed. Authors in \cite{shukla2011analysis} investigated if market power is one of the reasons for the price increases in the wholesale electricity market of India. Other regions have also been examined, including California \cite{borenstein1995market,wolak2003measuring}, England and Wales \cite{sweeting2007market}, and the Nordic wholesale markets \cite{hellmer2009evaluation}. The relationship between market power, congestion, and renewable energy has also been explored, such as in the Italian day-ahead market \cite{bigerna2016renewable}. Market power mitigation mechanisms in electricity markets have been proposed \cite{lin2022review}, including applying price caps as an efficient way to mitigate market power in terms of transmission congestion \cite{li2004market}. These efforts, among many others, demonstrate the importance of understanding if certain agents may exert market power in electricity systems. However, the market for SCC services is not included in these studies, leaving unexamined whether and how SGs at advantageous electrical positions would exploit market power for their own interests.

Regarding modeling techniques for capturing imperfect competition effects, bilevel optimization has proven to be a powerful tool. Through a bilevel model one can simulate a situation where a leader makes decisions before the follower, who then reacts to those decisions \cite{conejo2020complementarity}. Such a leader-follower structure, also known as a Mathematical Program with Equilibrium Constraints (MPEC), is well suited to capturing the interactions between strategic SGs and system operators in SCC service markets, providing a theoretical tool to fill the research gap described previously. Leaders, or price makers, are the agents with the capability to exert market power, due for example to their larger size in comparison to others, or their privileged location within a network. Mathematically, bilevel models are generally solved through replacing the lower-level problem (representing the follower) by its equivalent Karush-Kuhn-Tucker (KKT) conditions, which are then included as constraints in upper level, therefore leading to an equivalent single-level model \cite{conejo2020complementarity}. However, this approach strictly requires the lower-level problem to be continuous and convex, as otherwise the KKT conditions either do not exactly correspond to the original problem, or simply cannot be derived. Since SGs introduce discrete unit commitment variables, the traditional bilevel solution methods are not applicable in this work.

Several techniques for tackling non-convex bilevel models have been proposed. An iterative methodology based on column-and-constraint generation for the original problem \cite{zeng2014solving} has been widely employed. In particular, non-convex bilevel problems for energy markets are investigated in \cite{haghighat2018bilevel,jokar2022bilevel,zhang2017bilevel}, where  \cite{haghighat2018bilevel,jokar2022bilevel} address the challenge of commitment variables in UC, while \cite{zhang2017bilevel} tackles the binary variables that flag the installation of phase shifting transformers. Binary variables appear in both levels in \cite{pan2020bi} to characterize the interaction and on/off states of different objectives. Reference \cite{zhu2024optimal} also includes binary variables, in this case to consider the installation of certain equipment. However, this approach first used in \cite{zeng2014solving} would lead to a bilevel model independent of bids of strategic players, which contradicts our purpose of capturing potential strategic behaviors. An alternative method is to derive KKT conditions for each enumerated values of all binary variables \cite{huppmann2018exact}. This has the disadvantage of increasing the number of KKT conditions exponentially, especially when it comes to a large-scale problem with more combinations of binary states being considered, leading to high computational complexity. Therefore, the approach used in the present paper for solving the SCC-constrained bilevel UC problem is the one introduced in \cite{ye2019incorporating}. It is based on a primal-dual formulation which can tackle the non-convexity in the lower level, and being suitable for our purposes as the strategic bidding decisions for SCC may be considered as optimized variables.

\subsection{Research Flow and Main Contributions}
In this work, we start by describing SCC constraints and verifying their effectiveness in a linear form, laying the foundation for formulating the SCC market model. Then, we identify buses at risk of insufficient SCC and compute marginal prices for SCC services via the marginal unit pricing method proposed by \cite{chu2024pricing}, establishing a baseline to quantify perfect-competition market outcomes. Finally, the UC bilevel problem, which represents both energy and SCC service markets, is solved via a primal-dual formulation, enabling strategic analysis of how SGs at advantageous electrical positions exploit market power, and how such behavior impacts market clearing.

Specifically, the novel contributions of this work are:
\begin{enumerate}
    \item This paper presents a novel market formulation of SCC as a service using a bilevel optimization model, where the non-convexity arising from the binary commitment of SGs is handled through a primal–dual reformulation.
    \item The proposed model enables an analysis of imperfect competition in SCC service markets, showing that strategic SGs can earn up to three times higher revenues by raising ancillary service prices and operating for longer periods, which underscores the importance of careful market design.
\end{enumerate}

The remainder of this paper is structured as follows: Section \ref{SCC Constrained bilevel Problem Modeling} describes modeling of the SCC constraint and its application in the bilevel model. Section \ref{Solution Methodology} explains how to tackle the non-convexity of the model by primal-dual reformulation. Section \ref{Case Studies} includes relevant case studies to analyze market power issues and potential measures to mitigate such issues, while Section \ref{Conclusion} gives the conclusion and suggests lines for future work. Notably, the core work of this paper is the SCC‑constrained bilevel formulation with a tractable primal–dual restructuring.

\section{Short-Circuit Current Constrained Bilevel Problem Modeling}\label{SCC Constrained bilevel Problem Modeling}
This section introduces the SCC constraints that are next applied in a bilevel model to secure all buses to an acceptable level of SCC through a market setting. Note that only three-phase nodal short-circuit faults are considered. For the sake of clarity, time index `$t$' is omitted in the SCC expressions.

%\vspace{-5mm}
\subsection{Formulation of Short-Circuit Current Constraints}\label{Formulation of Short Circuit Current Constraints}
The SCC constraints derived in \cite{chu2021short} are used in this work, which incorporate the current contributions from both SGs and IBR. For a power system with bus $b \in \mathcal{B}$, SGs $g \in \mathcal{G}$ and IBR $c \in \mathcal{C}$, the SCC on bus $b$ can be expressed as:
\begin{equation}
     I_{b_\textrm{SC}}= \frac{\sum_{g\in\mathcal{G}}Z_{b\Psi(g)}\textrm{I}_gu_g+\sum_{c\in\mathcal{C}}Z_{b\Phi(c)}\textrm{I}_c\upalpha_c}{Z_{bb}}  \label{eq:define_of_SCC_constraints}  
\end{equation}
where \eqref{eq:define_of_SCC_constraints} accounts for the discrete nature of SCC from SGs through their commitment variable $u_g$. Furthermore, the SCC contribution is also dependent on the impedance $Z_{ij}$ and, for the case of IBR, their capacity factor $\upalpha_c$. $\textrm{I}_g$ and $\textrm{I}_c$ are the short-circuit injections from SGs and IBR, respectively.

The impedance matrix is calculated by inverting the corresponding admittance matrix, $Z_{ij}$ is thus hard to include in a duality-involved optimization problem. Therefore, a data-driven method is adopted to approximate the actual SCC constraints \cite{chu2021short}. Coefficients $\mathcal{K} = \{k_{b_g},k_{b_c},k_{b_m}\}$ are introduced to approximate the actual values of SCC \eqref{eq:define_of_SCC_constraints} into:
\begin{subequations} \label{eq:SCC_constraints_linearlized}
\begin{align}
& I_{b_\textrm{L}}= \sum_{g}k_{bg}u_{g}+\sum_{c}k_{bc}\upalpha_{c}+\sum_{m}k_{bm}\eta_{m} \ge \textrm{I}_{b_{\textrm{lim}}} \label{eq:define_SCC_constraints_linearized} \\
& \eta_m=u_{g_\textrm{1}}u_{g_\textrm{2}},\quad\textrm{s.t.}~\{g_\textrm{1},g_\textrm{2}\}=m    \label{eq:interactions_pair_SGs_1}         \\ 
& m\in\mathcal{M}=\{g_{\textrm{1}},g_{\textrm{2}}\mid \forall g_{\textrm{1}},\forall g_{\textrm{2}}\in\mathcal{G}\}    \label{eq:interactions_pair_SGs_2}    
    \end{align}
\end{subequations}
where \eqref{eq:define_SCC_constraints_linearized} is the approximate value of SCC, which is forced to be higher than $\textrm{I}_{b_\textrm{lim}}$. $\eta_m$ captures the interactions between pairs of SGs to simulate the nonlinearity, defined as~\eqref{eq:interactions_pair_SGs_1}-\eqref{eq:interactions_pair_SGs_2}. The approximate coefficients $\mathcal{K}$ are determined by operating an optimization-based classification procedure, which is detailed in Appendix~A. 

\subsection{Bilevel Model with Short-Circuit Current Constraints} \label{sec:bilevel_model}
The coefficients $\mathcal{K} = \{k_{b_g},k_{b_c},k_{b_m}\}$ in \eqref{eq:define_SCC_constraints_linearized} are constants in the bilevel model as discussed above, therefore \textit{italics} are no longer used for them. Besides, the number of dual variables corresponding to the auxiliary equations for linearizing the quadratic term $\eta_m$ increases exponentially with more SGs involved, leading to high computational complexity. Therefore, in the following models, we remove $\eta_m$ from the SCC constraint. Nevertheless, this simplifying assumption preserves the effectiveness of the formulation, as demonstrated in Section~\ref{Validation of Short Circuit Current Constraints}.

Note that the bilevel problem considered is an MPEC, that is, there is only one strategic player in the market who is able to manage multiple SGs, while all other participants are non-strategic players. Not simultaneously considering the strategic behaviors of all market participants allows us to capture the independent impact of this strategic player on the market and system operation, thereby decoupling the otherwise complex market problem. In fact, jointly considering multiple MPECs, corresponding to multiple strategic players participating in the market, would lead to an Equilibrium Problem with Equilibrium Constraints (EPEC). Interested readers are referred to \cite{conejo2020complementarity}.

\subsubsection{Upper-Level Model}
Strategic SGs maximize the profits from energy and SCC service markets at the Upper-Level (UL) problem, where the optimal bidding decisions are determined to form the bids reported to the lower level. The UL model is expressed as:
\begin{subequations}\label{eq:UL_model}
 \begingroup
\begin{alignat}{2}  
    &  \displaystyle \max_{ V_{\textrm{UL}}}   \sum_{t }  \Biggr[
     \sum_{\hat{g}} \biggl(\lambda^\textrm{E}_{t}  P_{\hat{g},t}  + \sum_{b } \lambda_{b,t}^{\textrm{SCC}}\textrm{k}_{b\hat{g}}u_{\hat{g},t} &&\nonumber \\
    & \qquad - \textrm{O}_{\hat{g}}^\textrm{nl}u_{\hat{g},t} -  \textrm{O}_{\hat{g}}^\textrm{m} P_{\hat{g},t}  - C^\textrm{st}_{\hat{g},t}  - C^\textrm{sh}_{\hat{g},t}  \biggl)  \Biggr]   &&
    \label{eq:UL_obj}
\end{alignat}
\endgroup
where:
\begin{alignat}{3}
 V_{\textrm{UL}}=\{\beta_{\hat{g},t}^\textrm{m}, \beta_{\hat{g},b,t}^\textrm{SCC}   \}   \label{eq:UL_cons}
 \end{alignat}
subject to:
\begin{alignat}{3}       
    & 1 \leq \beta_{\hat{g},t}^\textrm{m} \leq \overline{\upbeta}^\textrm{m}_{\hat{g}} \quad  \forall \hat{g}, \forall t     \label{eq:UL_cons_energy}\\
    & 1 \leq \beta_{\hat{g},b,t}^\textrm{SCC} \leq \overline{\upbeta}^\textrm{SCC}_{\hat{g},b} \quad \forall \hat{g}, \forall b, \forall t  \label{eq:UL_cons_AS} 
\end{alignat}
\end{subequations}
where objective function \eqref{eq:UL_obj} indicates that the revenues of strategic SGs come from the provision of energy and SCC, meanwhile the costs include no-load costs, marginal generation costs, start-up and shut-down costs. Eq.~\eqref{eq:UL_cons} shows the variables in UL problem. Eqs.~ \eqref{eq:UL_cons_energy}-\eqref{eq:UL_cons_AS} confine the strategic multipliers of bids.

\subsubsection{Lower-Level Model}
The system operator in the Lower-Level (LL) problem takes the strategic bids decided by \eqref{eq:UL_cons}, then minimizes the system-wide operation costs. The LL model is defined as:
\begin{subequations}\label{eq:LL_model}
 \begingroup
\begin{alignat}{2}
    &  \displaystyle \min_{V_\textrm{LL}}    \sum_{t}  \Biggr[
     \sum_{\hat{g} } \biggl(\textrm{O}_{\hat{g}}^\textrm{nl} u_{\hat{g},t} +\sum_{b}\textrm{O}_{\hat{g},b,t}^\textrm{SCC}\beta_{\hat{g},b,t}^\textrm{SCC} u_{\hat{g},t} +\textrm{O}_{\hat{g}}^\textrm{m} \beta_{\hat{g},t}^\textrm{m} P_{\hat{g},t} \nonumber \\ 
    & \qquad  +C_{\hat{g},t}^\textrm{st} + C_{\hat{g},t}^\textrm{sh}\biggl) +\sum_{\check{g} } \biggl(\textrm{O}_{\check{g}}^\textrm{nl} u_{\check{g},t} +\sum_{b}\textrm{O}_{\check{g},b,t}^\textrm{SCC} u_{\check{g},t} \nonumber \\
     & \qquad +\textrm{O}_{\check{g}}^\textrm{m} P_{\check{g},t}  + C_{\check{g},t}^\textrm{st} + C_{\check{g},t}^\textrm{sh} \biggl) +\sum_{c }\textrm{O}_c^{\textrm{l}} P_{c,t}    \Biggr]  \label{eq:LL_obj}  
\end{alignat}
\endgroup
where:
\begin{alignat}{3}
& V_\textrm{LL}  = \Bigl\{u_{\hat{g},t}, P_{\hat{g},t},C_{\hat{g},t}^\textrm{st},C_{\hat{g},t}^\textrm{sh}, \nonumber \\
& \hspace{0.2cm} \qquad \quad u_{\check{g},t}, P_{\check{g},t},C_{\check{g},t}^\textrm{st},C_{\check{g},t}^\textrm{sh}, P_{c,t} \Bigl\}     \label{eq:LL_variables} 
\end{alignat}
subject to:
\begin{align}  
    & \sum_{\hat{g}}\textrm{k}_{b\hat{g}}u_{\hat{g},t} + \sum_{\check{g}}\textrm{k}_{b\check{g}}u_{\check{g},t} \nonumber \\ 
    & +\sum_{c}\textrm{k}_{bc}\upalpha_{c,t}\ge \textrm{I}_{b_{\textrm{lim}}}:  (\lambda_{b,t}^{\textrm{SCC}}) \quad \forall b, \forall \hat{g}, \forall \check{g}, \forall c, \forall t \label{eq:LL_cons_SCC}  \\    
    & \textrm{P}^\textrm{D}_t = \hspace{-3pt}\sum_{\hat{g}}\hspace{-3pt}P_{\hat{g},t} + \sum_{\check{g} }\hspace{-3pt}P_{\check{g},t}+ \sum_{c}\hspace{-3pt}P_{c,t}:  (\lambda_{t}^{\textrm{E}}) \quad \forall \hat{g}, \forall \check{g}, \forall c, \forall t \label{eq:LL_cons_power_balance} \\[2pt]
    & u_{\hat{g},t} \textrm{P}_{\hat{g}}^\textrm{min}  \leq P_{\hat{g},t} \leq u_{\hat{g},t}  \textrm{P}_{\hat{g}}^\textrm{max}:  ({\mu}^{\textrm{min}}_{\hat{g},t}, {\mu}^{\textrm{max}}_{\hat{g},t}) \quad \forall \hat{g}, \forall t  \label{eq:LL_cons_strategic_SG_output}\\   
    & u_{\check{g},t} \textrm{P}_{\check{g}}^\textrm{min}  \leq P_{\check{g},t} \leq u_{\check{g},t}  \textrm{P}_{\check{g}}^\textrm{max}:  ({\mu}^{\textrm{min}}_{\check{g},t}, {\mu}^{\textrm{max}}_{\check{g},t}) \quad \forall \check{g}, \forall t  \label{eq:LL_cons_competitive_SG_output}\\
    & -\textrm{P}_{\hat{g}}^\textrm{rd}  \leq P_{\hat{g},t}-P_{\hat{g},(t-1)} \leq \textrm{P}_{\hat{g}}^\textrm{ru}:  ({\pi}^{\textrm{rd}}_{\hat{g},t}, {\pi}^{\textrm{ru}}_{\hat{g},t}) \quad \forall \hat{g}, \forall t  \label{eq:LL_cons_strategic_SG_ramp}\\
    & -\textrm{P}_{\check{g}}^\textrm{rd}  \leq P_{\check{g},t}-P_{\check{g},(t-1)} \leq \textrm{P}_{\check{g}}^\textrm{ru}:  ({\pi}^{\textrm{rd}}_{\check{g},t}, {\pi}^{\textrm{ru}}_{\check{g},t}) \quad \forall \check{g}, \forall t  \label{eq:LL_cons_competitive_SG_ramp}\\  
    & C_{\hat{g},t}^\textrm{st} \ge 0:  (\rho_{\hat{g},t}^\textrm{st}) \quad \forall \hat{g}, \forall t  \label{eq:LL_cons_stra_st_cost_positive}\\
    & C_{\hat{g},t}^\textrm{sh} \ge 0:  (\rho_{\hat{g},t}^\textrm{sh}) \quad \forall \hat{g}, \forall t  \label{eq:LL_cons_stra_sh_cost_positive}\\
    & C_{\check{g},t}^\textrm{st} \ge 0:  (\rho_{\check{g},t}^\textrm{st}) \quad \forall \check{g}, \forall t  \label{eq:LL_cons_comp_st_cost_positive}\\
    & C_{\check{g},t}^\textrm{sh} \ge 0:  (\rho_{\check{g},t}^\textrm{sh}) \quad \forall \check{g}, \forall t  \label{eq:LL_cons_comp_sh_cost_positive}\\ 
    & C_{\hat{g},t}^\textrm{st} \ge (u_{\hat{g},t}-u_{\hat{g},(t-1)})\textrm{K}^\textrm{st}_{\hat{g}}:  (\sigma_{\hat{g},t}^\textrm{st})\quad \forall \hat{g}, \forall t  \label{eq:LL_cons_st_stra_cost_lb}\\
    & C_{\hat{g},t}^\textrm{sh} \ge (u_{\hat{g},(t-1)}-u_{\hat{g},t})\textrm{K}^\textrm{sh}_{\hat{g}}:  (\sigma_{\hat{g},t}^\textrm{sh})\quad \forall \hat{g}, \forall t  \label{eq:LL_cons_sh_stra_cost_lb}\\ 
    & C_{\check{g},t}^\textrm{st} \ge (u_{\check{g},t}-u_{\check{g},(t-1)})\textrm{K}^\textrm{st}_{\check{g}}:  (\sigma_{\check{g},t}^\textrm{st})\quad \forall \check{g}, \forall t  \label{eq:LL_cons_st_comp_cost_lb}\\
    & C_{\check{g},t}^\textrm{sh} \ge (u_{\check{g},(t-1)}-u_{\check{g},t})\textrm{K}^\textrm{sh}_{\check{g}}:  (\sigma_{\check{g},t}^\textrm{sh})\quad \forall \check{g}, \forall t  \label{eq:LL_cons_sh_comp_cost_lb}\\
    & 0 \leq P_{c,t} \leq \upalpha_{c,t}\textrm{P}^{\textrm{max}}_{c}:  (\zeta_{c,t}^\textrm{min},\zeta_{c,t}^\textrm{max})\quad \forall c, \forall t  \label{eq:LL_cons_wt}\\
    & u_{\hat{g},t},u_{\check{g},t} \in \{0,1\} \quad \forall \hat{g}, \forall \check{g}, \forall t \label{eq:LL_binary_SGs}
\end{align}   
\end{subequations}  
where the system operating costs are represented in \eqref{eq:LL_obj}, including the costs explained in \eqref{eq:UL_obj}, the costs of SCC service and energy provided by IBR. Eq.~\eqref{eq:LL_variables} includes primal variables in the LL model which follow constraints: SCC on bus $b$ contributed by SGs and IBR \eqref{eq:LL_cons_SCC}; Demand-supply power balance constraints~\eqref{eq:LL_cons_power_balance}; Minimum stable generation and maximum generation constraints for SGs \eqref{eq:LL_cons_strategic_SG_output}-\eqref{eq:LL_cons_competitive_SG_output}; Ramp-up/down constraints \eqref{eq:LL_cons_strategic_SG_ramp}-\eqref{eq:LL_cons_competitive_SG_ramp}; Start-up/shut-down costs \eqref{eq:LL_cons_stra_st_cost_positive}-\eqref{eq:LL_cons_sh_comp_cost_lb}; Minimum and maximum generation constraints for IBR \eqref{eq:LL_cons_wt}; Eq.~\eqref{eq:LL_binary_SGs} imposes integrality for commitment variables of SGs. The SCC service offer $\textrm{O}_{g,b,t}^\textrm{SCC}$ is determined by the marginal unit pricing, which is elaborated in Appendix~B.

\section{Solution Methodology}\label{Solution Methodology}
Given that the bilevel model in this work contains binary variables in the lower level through \eqref{eq:LL_binary_SGs}, converting the LL model into the equivalent KKT conditions becomes impractical. A primal-dual formulation is adopted to tackle this non-convex bilevel model, which is designed to restate the LL problem by deriving the dual with relaxed binary variables. Then, after recovering the discreteness of commitment decisions, the bilevel problem can be further converted to a single-level modeling by minimizing the duality gap between the primal and dual, as we discuss next.

\subsection{Dualization of Primal Lower-Level Problem}
The binary variables are relaxed to be continuous within a limited range, indicated as follows:
% \vspace{-0.5cm}
\begin{subequations}\label{eq:Dual_relaxed_binary}
\begin{alignat}{2}
0\leq u_{\hat{g},t}\leq 1:(\psi_{\hat{g},t}^{\textrm{min}},\psi_{\hat{g},t}^{\textrm{max}}) \quad  \forall \hat{g}, \forall t  \\
0\leq u_{\check{g},t}\leq 1: (\psi_{\check{g},t}^{\textrm{min}},\psi_{\check{g},t}^{\textrm{max}}) \quad  \forall \check{g}, \forall t 
\end{alignat}
\end{subequations}
with this relaxation, the dual variables for all constraints in \eqref{eq:LL_model} can then be defined. Consequently, the primal LL problem is now defined by \eqref{eq:LL_obj}-\eqref{eq:LL_cons_wt} and \eqref{eq:Dual_relaxed_binary}.

After the relaxation, the dual problem associated with the relaxed LL problem is derived as follows:
\begin{subequations}\label{eq:dual_LL_model}
\begin{alignat}{2}
    &  \displaystyle \max_{V_\textrm{DLL}}    \sum_{t}  \Biggr[\textrm{P}^\textrm{D}_t\lambda_t^{\textrm{E}}+\sum_b\textrm{I}_{b_{\textrm{lim}}}\lambda_{b,t}^{\text{SCC}} - \hspace{-0.25cm} \sum_c\upalpha_{c,t}\textrm{P}^{\textrm{max}}_{c}\zeta_{c,t}^\textrm{max}- \hspace{-0.25cm} \sum_{\hat{g}}\psi_{\hat{g},t}^{\textrm{max}} \nonumber \\  
    & -\sum_{\check{g}}\psi_{\check{g},t}^{\textrm{max}} \Biggr]+ \sum_{\hat{g},(t=1)} \hspace{-0.25cm} P_{\hat{g},0}\pi_{\hat{g},t}^{\textrm{rd}}-\sum_{\hat{g},t}\textrm{P}_{\hat{g}}^{\textrm{rd}}\pi_{\hat{g},t}^{\textrm{rd}}+\sum_{\check{g},(t=1)} \hspace{-0.25cm} P_{\check{g},0}\pi_{\check{g},t}^{\textrm{rd}} \nonumber \\ 
    & -\sum_{\check{g},t}\textrm{P}_{\check{g}}^{\textrm{rd}}\pi_{\check{g},t}^{\textrm{rd}}-\sum_{\hat{g},(t=1)} \hspace{-0.25cm} P_{\hat{g},0}\pi_{\hat{g},t}^{\textrm{ru}}-\sum_{\hat{g},t}\textrm{P}_{\hat{g}}^{\textrm{ru}}\pi_{\hat{g},t}^{\textrm{ru}}-\sum_{\check{g},(t=1)} \hspace{-0.25cm} P_{\check{g},0}\pi_{\check{g},t}^{\textrm{ru}}  \nonumber \\
    & -\sum_{\check{g},t}\textrm{P}_{\check{g}}^{\textrm{ru}}\pi_{\check{g},t}^{\textrm{ru}}-\sum_{\hat{g},(t=1)} \hspace{-0.25cm} u_{\hat{g},0}\textrm{K}_{\hat{g}}^{\textrm{st}}\sigma_{\hat{g},t}^{\textrm{st}}-\sum_{\check{g},(t=1)} \hspace{-0.25cm} u_{\check{g},0}\textrm{K}_{\check{g}}^{\textrm{st}}\sigma_{\check{g},t}^{\textrm{st}} \nonumber \\
    & +\sum_{\hat{g},(t=1)} \hspace{-0.25cm} u_{\hat{g},0}\textrm{K}_{\hat{g}}^{\textrm{sh}}\sigma_{\hat{g},t}^{\textrm{sh}}+\sum_{\check{g},(t=1)} \hspace{-0.25cm} u_{\check{g},0}\textrm{K}_{\check{g}}^{\textrm{sh}}\sigma_{\check{g},t}^{\textrm{sh}}     
    \label{eq:DLL_obj}  
\end{alignat}
where:
\begin{alignat}{2}
& V_\textrm{DLL}  = \Bigl\{ \lambda_t^{\textrm{E}},\lambda_{b,t}^{\textrm{SCC}},\zeta_{c,t}^\textrm{max},\nonumber \\ 
& \qquad \qquad \mu^{\textrm{min}}_{\hat{g},t}, {\mu}^{\textrm{max}}_{\hat{g},t}, {\pi}^{\textrm{rd}}_{\hat{g},t}, {\pi}^{\textrm{ru}}_{\hat{g},t},\sigma_{\hat{g},t}^\textrm{st},\sigma_{\hat{g},t}^\textrm{sh},\psi_{\hat{g},t}^{\textrm{max}}, \nonumber \\ 
& \qquad \qquad \mu^{\textrm{min}}_{\check{g},t}, {\mu}^{\textrm{max}}_{\check{g},t}, {\pi}^{\textrm{rd}}_{\check{g},t}, {\pi}^{\textrm{ru}}_{\check{g},t},\sigma_{\check{g},t}^\textrm{st},\sigma_{\check{g},t}^\textrm{sh},\psi_{\check{g},t}^{\textrm{max}}\Bigl\}    \label{eq:DLL_variables} 
\end{alignat}
subject to:
\begin{alignat}{2}
    & \textrm{O}_{\hat{g}}^\textrm{nl} +\sum_{b}\beta_{\hat{g},b,t}^\textrm{SCC}\textrm{O}_{\hat{g},b,t}^\textrm{SCC}-\textrm{k}_{b \hat{g}}\lambda_{b,t}^{\textrm{SCC}}-\textrm{P}_{\hat{g}}^{\textrm{max}}\mu_{\hat{g},t}^{\textrm{max}}+\textrm{P}_{\hat{g}}^{\textrm{min}}\mu_{\hat{g},t}^{\textrm{min}} \nonumber \\
    &+\textrm{K}_{\hat{g}}^{\textrm{st}}(\sigma_{\hat{g},t}^{\textrm{st}}-\sigma_{\hat{g},(t+1)}^{\textrm{st}})+\textrm{K}_{\hat{g}}^{\textrm{sh}}(\sigma_{\hat{g},(t+1)}^{\textrm{sh}}-\sigma_{\hat{g},t}^{\textrm{sh}}) \nonumber \\ 
    &+\psi_{\hat{g},t}^{\textrm{max}} \ge 0, \quad \forall \hat{g}, \forall b, \forall t \leq T-1  \label{eq:DLL_cons_binary_stra_morethan1} \\[3pt]
    & \textrm{O}_{\hat{g}}^\textrm{nl} +\sum_{b}\beta_{\hat{g},b,t}^\textrm{SCC}\textrm{O}_{\hat{g},b,t}^\textrm{SCC} -\textrm{k}_{b \hat{g}}\lambda_{b,t}^{\textrm{SCC}}-\textrm{P}_{\hat{g}}^{\textrm{max}}\mu_{\hat{g},t}^{\textrm{max}}+\textrm{P}_{\hat{g}}^{\textrm{min}}\mu_{\hat{g},t}^{\textrm{min}} \nonumber \\
    &+\textrm{K}_{\hat{g}}^{\textrm{st}}\sigma_{\hat{g},t}^{\textrm{st}}-\textrm{K}_{\hat{g}}^{\textrm{sh}}\sigma_{\hat{g},t}^{\textrm{sh}}+\psi_{\hat{g},t}^{\textrm{max}} \ge 0,  \quad \forall \hat{g}, \forall b, \forall t = T  \label{eq:DLL_cons_binary_stra_only1} \\[3pt]
    & \textrm{O}_{\check{g}}^\textrm{nl} +\sum_{b}\textrm{O}_{\check{g},b,t}^\textrm{SCC} -\textrm{k}_{b \check{g}}\lambda_{b,t}^{\textrm{SCC}}-\textrm{P}_{\check{g}}^{\textrm{max}}\mu_{\check{g},t}^{\textrm{max}}+\textrm{P}_{\check{g}}^{\textrm{min}}\mu_{\check{g},t}^{\textrm{min}} \nonumber \\
    &+\textrm{K}_{\check{g}}^{\textrm{st}}(\sigma_{\check{g},t}^{\textrm{st}}-\sigma_{\check{g},(t+1)}^{\textrm{st}})+\textrm{K}_{\check{g}}^{\textrm{sh}}(\sigma_{\check{g},(t+1)}^{\textrm{sh}}-\sigma_{\check{g},t}^{\textrm{sh}}) \nonumber \\ 
    &+\psi_{\check{g},t}^{\textrm{max}} \ge 0, \quad  \forall \check{g}, \forall b, \forall t \leq T-1  \label{eq:DLL_cons_binary_comp_morethan1} \\[3pt]
    & \textrm{O}_{\check{g}}^\textrm{nl} +\sum_{b}\textrm{O}_{\check{g},b,t}^\textrm{SCC} -\textrm{k}_{b \check{g}}\lambda_{b,t}^{\textrm{SCC}}-\textrm{P}_{\check{g}}^{\textrm{max}}\mu_{\check{g},t}^{\textrm{max}}+\textrm{P}_{\check{g}}^{\textrm{min}}\mu_{\check{g},t}^{\textrm{min}} \nonumber \\
    &+\textrm{K}_{\check{g}}^{\textrm{st}}\sigma_{\check{g},t}^{\textrm{st}}-\textrm{K}_{\check{g}}^{\textrm{sh}}\sigma_{\check{g},t}^{\textrm{sh}}+\psi_{\check{g},t}^{\textrm{max}} \ge 0,  \quad \forall \check{g}, \forall b, \forall t = T  \label{eq:DLL_cons_binary_comp_only1} \\[3pt]
    & \textrm{O}_{\hat{g}}^\textrm{m}b_{\hat{g},t}^\textrm{m}-\lambda_{t}^{\textrm{E}}+\mu^{\textrm{max}}_{\hat{g},t}-\mu^{\textrm{min}}_{\hat{g},t}+\pi^{\textrm{ru}}_{\hat{g},t}-\pi^{\textrm{ru}}_{\hat{g},(t+1)} \nonumber \\
    & -\pi^{\textrm{rd}}_{\hat{g},t}+\pi^{\textrm{rd}}_{\hat{g},(t+1)} \ge 0,  \quad \forall \hat{g}, \forall t \leq T-1  \label{eq:DLL_cons_Pg_morethan1} \\[3pt]
    & \textrm{O}_{\hat{g}}^\textrm{m}b_{\hat{g},t}^\textrm{m}-\lambda_{t}^{\textrm{E}}+\mu^{\textrm{max}}_{\hat{g},t}-\mu^{\textrm{min}}_{\hat{g},t}+\pi^{\textrm{ru}}_{\hat{g},t}-\pi^{\textrm{rd}}_{\hat{g},t} \ge 0,  \nonumber \\
    & \quad \forall \hat{g}, \forall t = T  \label{eq:DLL_cons_Pg_only1} \\[3pt]
    & \textrm{O}_{\check{g}}^\textrm{m}-\lambda_{t}^{\textrm{E}}+\mu^{\textrm{max}}_{\check{g},t}-\mu^{\textrm{min}}_{\check{g},t}+\pi^{\textrm{ru}}_{\check{g},t}-\pi^{\textrm{ru}}_{\check{g},(t+1)} \nonumber \\
    & -\pi^{\textrm{rd}}_{\check{g},t}+\pi^{\textrm{rd}}_{\check{g},(t+1)} \ge 0,  \quad \forall {\check{g}}, \forall t \leq T-1  \label{eq:DLL_cons_comp_Pg_morethan1} \\[3pt]
    & \textrm{O}_{\check{g}}^\textrm{m}-\lambda_{t}^{\textrm{E}}+\mu^{\textrm{max}}_{\check{g},t}-\mu^{\textrm{min}}_{\check{g},t}+\pi^{\textrm{ru}}_{\check{g},t}-\pi^{\textrm{rd}}_{\check{g},t} \ge 0,  \nonumber \\
    & \quad \forall \check{g}, \forall t = T  \label{eq:DLL_cons_comp_Pg_only1} \\[3pt]
    & 1-\sigma_{\hat{g},t}^{\textrm{st}} \ge 0,  \quad \forall \hat{g}, \forall t \label{eq:DLL_cons_stra_Cst} \\[3pt]
    & 1-\sigma_{\hat{g},t}^{\textrm{sh}} \ge 0,  \quad \forall \hat{g}, \forall t \label{eq:DLL_cons_stra_Csh} \\[3pt]
    & 1-\sigma_{\check{g},t}^{\textrm{st}} \ge 0,  \quad \forall \check{g}, \forall t \label{eq:DLL_cons_comp_Cst} \\[3pt]
    & 1-\sigma_{\check{g},t}^{\textrm{sh}} \ge 0,  \quad \forall \check{g}, \forall t \label{eq:DLL_cons_comp_Csh} \\[3pt]
    & \textrm{O}_{c}^\textrm{l}-\lambda_t^{\textrm{E}}+\zeta^{\textrm{max}}_{c,t} \ge 0,  \quad \forall c, \forall t \label{eq:DLL_cons_Pc}\\[3pt]
   % &  \lambda_{F,t}-\omega^{\textrm{lim}}_{F,t}\ge 0,  \quad \forall F, \forall t \label{eq:DLL_cons_SCC_lim}\\[3pt]      
    & V_\textrm{DLL} \in \mathbb{R}_+,  \quad \forall b, \forall \hat{g}, \forall \check{g}, \forall c, \forall t \label{eq:DLL_var_nonnegative}
\end{alignat}
\end{subequations}
where \eqref{eq:DLL_obj} is the objective function of the dual problem. Eq.~\eqref{eq:DLL_variables} shows the dual variables involved. Eqs.~\eqref{eq:DLL_cons_binary_stra_morethan1}-\eqref{eq:DLL_cons_binary_comp_only1}, \eqref{eq:DLL_cons_Pg_morethan1}-\eqref{eq:DLL_cons_comp_Pg_only1}, \eqref{eq:DLL_cons_stra_Cst}-\eqref{eq:DLL_cons_comp_Csh} and \eqref{eq:DLL_cons_Pc} represent dual constraints regarding primal variables: $u_{\hat{g},t}$,$u_{\check{g},t}$; $P_{\hat{g},t}$, $P_{\check{g},t}$; $C_{\hat{g},t}^\textrm{st}$,$C_{\hat{g},t}^\textrm{sh}$,$C_{\check{g},t}^\textrm{st}$,$C_{\check{g},t}^\textrm{sh}$ and $P_{c,t}$, respectively. Eq.~\eqref{eq:DLL_var_nonnegative} expresses non-negativity of the dual variables. Note that $\lambda_t^{\textrm{E}}$ is also lower bounded by zero here to ensure non-negative energy revenues for SGs, otherwise they would not comply with the dispatch due to insufficient financial incentives. Meanwhile, it provides a reasonable lower bound for linearizing the bilinear term, as shown in the Appendix A.

\subsection{Primal-Dual Single-Level Optimization Model}
Binary variables are only relaxed for deriving the dual problem. By recovering~\eqref{eq:LL_binary_SGs}, they are constrained to be integer again to capture the UC property. This implies that the reformulated LL model is not completely equivalent to the primal, thus holding weak duality. The penalty function method with a parameter W can be employed to minimize the Duality Gap (DG) to produce feasible solutions, as introduced in \cite{ye2019incorporating}. Therefore, the final form of the problem is:
% \vspace{-0.1cm}
\begin{subequations}\label{eq:primal_dual_model}
\begin{alignat}{2}
    &  \displaystyle \max_{V}  \eqref{eq:UL_obj}-\textrm{W} \cdot DG \label{eq:pri_dual_obj}
\end{alignat}
% \vspace{-0.2cm}
where: 
\begin{alignat}{2}
    &  DG = \eqref{eq:LL_obj}-\eqref{eq:DLL_obj}   \label{eq:pri_dual_DG}  \\
    & V = \{ V_\textrm{UL}, V_{\textrm{LL}}, V_{\textrm{DLL}} \label{eq:pri_dual_V} \}
\end{alignat}
subject to:
\begin{alignat}{3}
    &  \eqref{eq:UL_cons_energy}\textrm{-}\eqref{eq:UL_cons_AS}, \eqref{eq:LL_cons_SCC}\textrm{-}\eqref{eq:LL_binary_SGs}, \eqref{eq:DLL_cons_binary_stra_morethan1}\textrm{-}\eqref{eq:DLL_var_nonnegative} \label{eq:pri_dual_cons}
\end{alignat}
\end{subequations}  
where \eqref{eq:primal_dual_model} is a single-level problem. The objective function \eqref{eq:pri_dual_obj} couples the maximization of UL objective \eqref{eq:UL_obj} and minimization of DG between primal and dual LL problems. The DG is defined as \eqref{eq:pri_dual_DG}. Eqs.~\eqref{eq:pri_dual_V}-\eqref{eq:pri_dual_cons} clarify the variables and constraints involved. The positive penalty factor W governs the trade-offs of objectives, i.e., the accuracy of market clearing and the consideration of profits. An appropriate value for this parameter must be found for the problem at hand, since a value too small for W would not sufficiently penalize the DG, while a value too large would instead drive the algorithm to overly focus on the DG minimization, and render the profit of strategic agents in \eqref{eq:primal_dual_model} less important, potentially leading to less profitable strategic behaviors, as demonstrated in Section~\ref{Two Strategic SGs in Single bus}. Therefore, we define a ratio associated with primal and dual LL problems, to evaluate the error of solutions as follows:
\begin{equation}\label{eq:ratio_DG}
     r^{\textrm{DG}}= \frac{\eqref{eq:pri_dual_DG}} {\eqref{eq:LL_obj}}
 \end{equation}
where a small value of $r^{\textrm{DG}}$ indicates a minor gap between the primal and dual problems, i.e., a more accurate solution of market clearing.

It is noteworthy that problem \eqref{eq:primal_dual_model} is still nonlinear, as it still contains quadratic terms $\lambda^\textrm{E}_{t}  P_{\hat{g},t}$, $\lambda_{b,t}^{\textrm{SCC}}\textrm{k}_{b\hat{g}}u_{\hat{g},t}$ in \eqref{eq:UL_obj}, $\textrm{O}_{\hat{g},b,t}^\textrm{SCC}\beta_{\hat{g},b,t}^\textrm{SCC} u_{\hat{g},t}$, $\textrm{O}_{\hat{g}}^\textrm{m} \beta_{\hat{g},t}^\textrm{m} P_{\hat{g},t}$ in \eqref{eq:LL_obj}. These bilinear terms can be linearized using McCormick envelopes with appropriate bounds on the associated variables, as demonstrated in Appendix~C.

\begin{figure}[t]
\centering
\includegraphics[width=1\columnwidth]{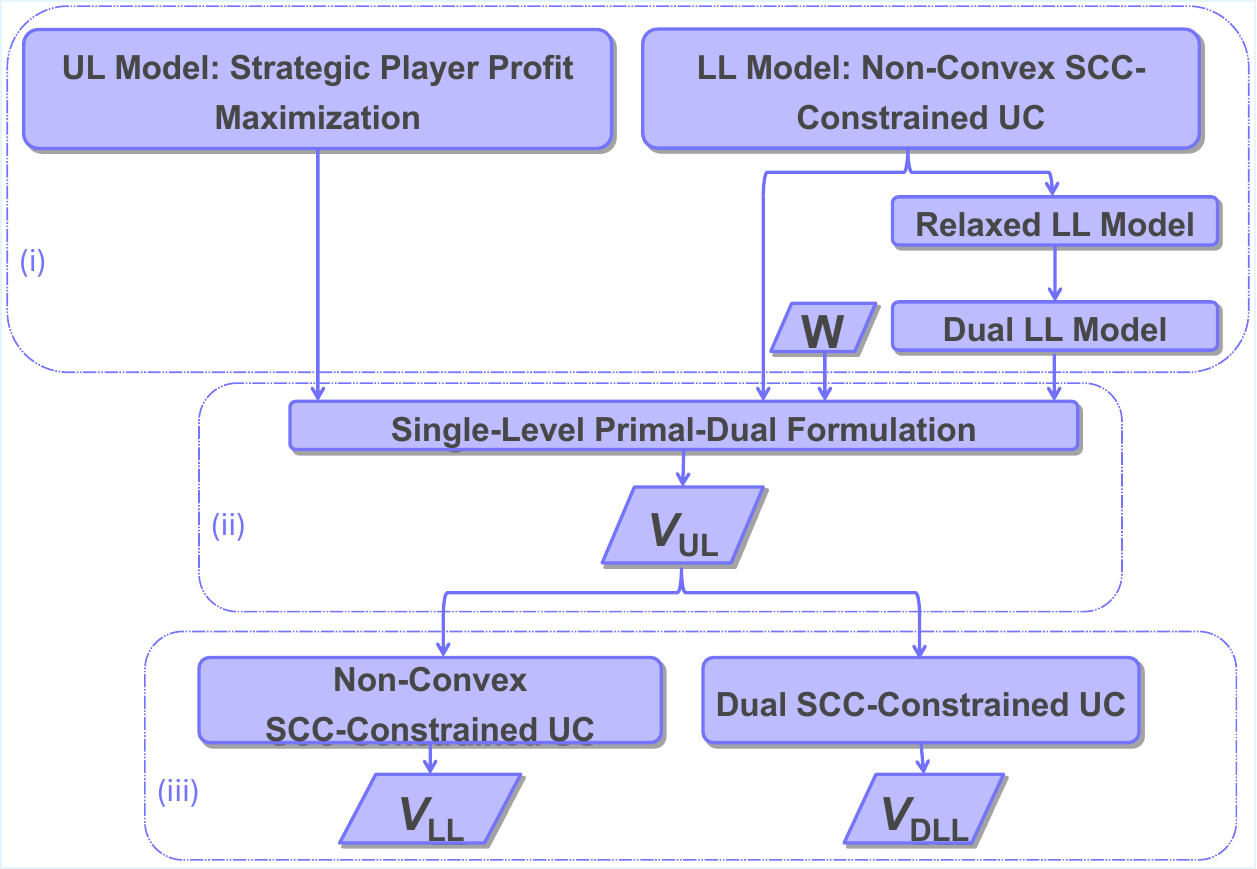}
\caption{Diagram of solution methodology.}
\label{Diagram of solution methodology}
     %\vspace{-0.2cm}
\end{figure}

In summary, to solve the nonlinear bilevel model with non-convexity, binary variables are relaxed to be continuous to derive the dual of the primal LL problem, as seen in block (i) in Fig.~\ref{Diagram of solution methodology}. Then they are tightened to achieve realistic UC solutions, although this leads to a DG between the primal and dual LL problems. The penalty function method with parameter W is used to minimize the DG, giving a trade-off between the solution accuracy and values of objectives (block (ii)). Finally, nonlinear terms are restated to form the final single-level, mixed-integer linear program, in which the primal and dual LL variables are co-optimized with UL variables (block (iii)).

\section{Case Studies}\label{Case Studies}
Here, the basic system setup is first described (Section~\ref{Test System Setting}) and the effectiveness of the SCC constraints is validated (Section~\ref{Validation of Short Circuit Current Constraints}). In the formal market power analysis, we begin by examining a perfectly competitive market case (Section~\ref{System Operation in the Perfectly Competitive Market}), which serves as a basis for comparison with subsequent imperfectly competitive cases (Section~\ref{System Operation in the Imperfectly Competitive Market}). We then analyze the market outcomes under different strategic ownership structures, selected to reflect critical electrical locations whose unique SCC provision would shape potential market power, including two strategic SGs located at the same node (Section~\ref{Two Strategic SGs in Single bus}) and cases with SGs located at different nodes (Section~\ref{A Strategic Player Managing SGs in Electrically Distanced buses}). In addition, we investigate possible measures to mitigate the adverse effects of market power (Section~\ref{Potential Countermeasures to Mitigate Market Power}), and finally discuss how the adopted system and market settings may influence the analysis presented here (Section~\ref{Potential Impact of System Operation on Market Power in SCC Services}).

\subsection{Test System Setting}\label{Test System Setting}

\begin{figure}[t]
\centering
\includegraphics[width=0.8\columnwidth]{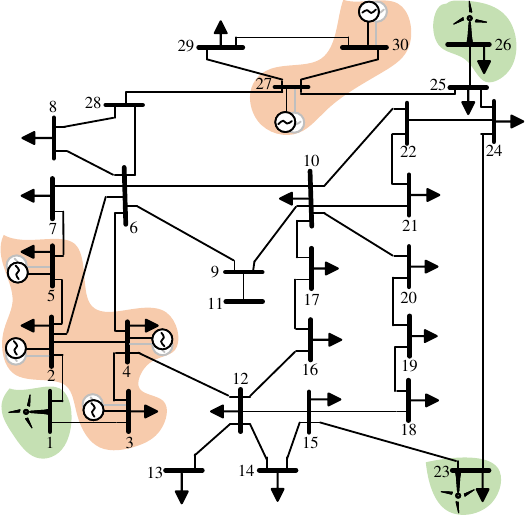}
\caption{Modified IEEE 30-bus system.}
\label{ieee_system}
     %\vspace{-0.2cm}
\end{figure}

\begin{table}[t]
\centering
\caption{Operating Parameters of Synchronous Generators}
%\vspace{-0.2cm}
\setlength{\tabcolsep}{5pt}
{\fontsize{8pt}{12pt}\selectfont
\begin{tabular}{lcccccc}
\toprule
Bus & 2 & 3 & 4 & 5 & 27 & 30 \\
\midrule
\(\textrm{O}_{g}^\textrm{nl}\) (\texteuro/h) & 1,743 & 1,501 & 1,376 & 1,093 & 990 & 857 \\
\(\textrm{O}_{g_1}^\textrm{m}\) (\texteuro/MWh) & 6.20 & 7.10 & 10.47 & 12.28 & 13.53 & 15.36 \\
\(\textrm{O}_{g_2}^\textrm{m}\) (\texteuro/MWh) & 7.07 & 8.72 & 11.49 & 12.84 & 14.60 & 15.02 \\
\(\textrm{K}_g^\textrm{st}\) (\texteuro/h) & 20,000 & 12,500 & 9,250 & 7,200 & 5,500 & 3,100 \\
\(\textrm{K}_g^\textrm{sh}\) (\texteuro/h) & 5,000 & 2,850 & 1,850 & 1,440 & 1,200 & 1,000 \\
\(\textrm{P}_g^\textrm{min}\) (MW) & 658 & 576 & 302 & 133 & 130 & 58 \\
\(\textrm{P}_g^\textrm{max}\) (MW) & 1,317 & 1,152 & 756 & 667 & 650 & 576 \\
\(\textrm{P}_g^\textrm{rd}\) (MW/h) & 263 & 230 & 302 & 267 & 390 & 346 \\
\(\textrm{P}_g^\textrm{ru}\) (MW/h) & 263 & 230 & 302 & 267 & 390 & 346 \\
\(u_{g,0}\) & 1 & 1 & 1 & 1 & 1 & 0 \\
\(P_{g,0}\) (MW) & 1,054 & 922 & 605 & 534 & 520 & 0 \\
\bottomrule
\end{tabular}
}
\label{table:SGs_para}
%\vspace{-0.3cm}
\end{table}

\begin{table}[t]
\centering
\caption{Bus Locations of Generators}
\setlength{\tabcolsep}{5pt}
{\fontsize{8pt}{12pt}\selectfont
\begin{tabular}{lcc}
\toprule
Generator type & IBR & SGs \\
\midrule
\qquad  Bus & 1, 23, 26 & 2, 3, 4, 5, 27, 30 (2 SGs/bus) \\
\bottomrule
\end{tabular}
}
\label{table:resource_bus}
\end{table}

To examine whether market power issues may arise in SCC services, certain units under an appropriate system scale should be selected, a modified IEEE 30-bus system (as seen in Fig.~\ref{ieee_system}) is therefore adopted in the case studies. The IBR, i.e., wind turbines are added to buses \{1, 23, 26\}, with energy bids \{2.46, 3.19, 2.86\} \texteuro/MWh, respectively. In order to model a modular SCC contribution from SGs, we consider that two units are located in each bus containing an SG, i.e., buses \{2, 3, 4, 5, 27, 30\}. This way, diverse decisions can be made in the market. For example, different UC status (i.e., both online, both offline, just one of them online) can determine various SCC contributions and bids. The SGs in bus 2 are separately named as $g_1$-$b$2 and $g_2$-$b$2, with 2$g$-$b$2 standing for both of them. Other thermal units follow the same nomenclature. The SGs' parameters are given in Table \ref{table:SGs_para}. The locations of units are summarized in Table \ref{table:resource_bus}. The selected system exhibits typical weak-grid characteristics at remote IBR buses and strong topological asymmetry in SCC contributions. For other given systems, such as larger-scale ones, the proposed methodology could be applied to analyze such market power concerns if they indeed exist.

The network impedances and reactances of SGs refer to \cite{shahidehpour2004communication} and \cite{reichert1974computation}, respectively. Load demand $\textrm{P}^\textrm{D}_t\in [\textrm{5109}, \textrm{7689}]$ MW. Total installed wind generation capacity is 750 MW. Base power is $S_B=\textrm{100}~\textrm{MVA}$. $\textrm{I}_c$, the short-circuit injection from IBR, is 1.00 p.u. Nominal voltage is set as 0.95 p.u. for conservativeness \cite{chu2021short}. Parameter $\nu$ is set as 0.1 p.u. to yield a desired approximation for SCC constraints. Maximum multipliers for bidding decisions $\overline{\upbeta}^\textrm{m}_{\hat{g}}$ and $\overline{\upbeta}^\textrm{SCC}_{\hat{g},b}$ are both set as 2. All code and data are available at the public repository \cite{Code}. All key parameters used in the case study are listed in Appendix~D.

The simulations in this work were processed by \texttt{Julia-JuMP} and \texttt{Gurobi} in version 12.0.0. The Gurobi solver is employed with default settings, where the key parameters are set as follows: the relative gap tolerance (MIPGap) is 1e-4, the feasibility tolerance (FeasibilityTol) and optimality tolerance (OptimalityTol) are both 1e-6, and the integer feasibility tolerance (IntFeasTol) is 1e-5. Other solver parameters (e.g., threads, presolve) use the solver’s auto-selected defaults.

\subsection{Validation of Short-Circuit Current Constraints}\label{Validation of Short Circuit Current Constraints}
The approximate SCC constraints \eqref{eq:define_SCC_constraints_linearized} are linearized to be \eqref{eq:LL_cons_SCC} for the primal-dual formulation. Two evaluations are first conducted to validate the obtained SCC constraints.

\textit{\textbf{Evaluation 1}}: Training errors of approximation. Two types of errors are analyzed to verify the approximation \cite{chu2024pricing}: Type-I: $I_{b_{\textrm{L}}}\geq \textrm{I}_{b_{\textrm{lim}}} \cap I_{{b_{\textrm{SC}}}}<\textrm{I}_{b_{\textrm{lim}}}$ means the approximate SCC is looser than the actual requirement, a potentially dangerous situation; Type-II: $I_{b_{\textrm{L}}}<\textrm{I}_{b_{\textrm{lim}}}\cap I_{{b_{\textrm{SC}}}}\geq \textrm{I}_{b_{\textrm{lim}}}$ implies that the approximate SCC defines a tighter boundary than the grid-code value $\textrm{I}_{b_{\textrm{lim}}}$, therefore satisfying system security. Type-I errors are likely to cause insecure operating conditions, thus being undesired. The averaged value of the errors is defined below:
\begin{equation}\label{eq:error}
 N_e=| \xi |,~   err = \frac{1}{N_e} \sum_{\xi} \frac{I^{(\xi)}_{b_\textrm{L}} - I^{(\xi)}_{b_\textrm{SC}}}{I^{(\xi)}_{b_\textrm{SC}}}
\end{equation}
where $\xi$ indicates the enumerated system operating points. $N_e$ is the total number of errors.

\textit{\textbf{Evaluation 2}}: Minimum SCC distribution. We consider all buses to comply with approximate and linear SCC constraints respectively in an economic dispatch problem. The minimum SCC of all buses can then be calculated to verify whether the SCC at each bus and at each time period remains above $\textrm{I}_{b_{\textrm{lim}}}$. 

\begin{table}[!t]
\centering
\caption{Assessment of Minimum SCC and Training Errors for Different SCC Requirements}
{\fontsize{8pt}{12pt}\selectfont
\setlength{\tabcolsep}{2pt}
\begin{tabular}{c
                cc
                cc
                cc
                cc}
\toprule
\multirow{3}{*}{\(\textrm{I}_{b_{\textrm{lim}}}\) (p.u.)} 
& \multicolumn{4}{c}{Minimum SCC} 
& \multicolumn{2}{c}{Type-I} 
& \multicolumn{2}{c}{Type-II} \\
\cmidrule(lr){2-5} \cmidrule(lr){6-7} \cmidrule(lr){8-9}
& \multicolumn{2}{c}{Approximate \eqref{eq:define_SCC_constraints_linearized}} 
& \multicolumn{2}{c}{Linearized \eqref{eq:LL_cons_SCC}} 
& \multirow{2}{*}{\(N_e\)} 
& \multirow{2}{*}{$err$} 
& \multirow{2}{*}{\(N_e\)} 
& \multirow{2}{*}{$err$} \\
\cmidrule(lr){2-3} \cmidrule(lr){4-5}
& p.u. & Bus & p.u. & Bus & & & & \\
\midrule
3.00 & 4.50 & 30 & 4.73 & 30 & 0 & 0 & 2 & -1.13\% \\
4.00 & 4.13 & 30 & 4.06 & 30 & 0 & 0 & 8 & -0.56\% \\
5.00 & 5.04 & 26 & 5.20 & 30 & 0 & 0 & 11 & -1.79\% \\
\bottomrule
\end{tabular}
}
\label{table:SCC_validation}
%\vspace{-0.4cm}
\end{table}

Results evaluated at different $\textrm{I}_{b_{\textrm{lim}}}$ are presented in Table \ref{table:SCC_validation}. It is shown that the approximation eliminates Type-I errors and produces only a small amount of Type-II errors, indicating a good performance. The minimum SCC of the entire system is above the $\textrm{I}_{b_{\textrm{lim}}}$, meaning that the SCC level on all buses is always within the safe range, justifying the approximate SCC constraints \eqref{eq:define_SCC_constraints_linearized}. The effectiveness of the linearized SCC constraint \eqref{eq:LL_cons_SCC} is demonstrated by the results on the middle section of Table~\ref{table:SCC_validation}, showing that minimum SCC is still beyond $\textrm{I}_{b_{\textrm{lim}}}$. 

\subsection{System Operation under Perfect Competition}\label{System Operation in the Perfectly Competitive Market}
Here, we examine a perfectly competitive case with a complete operation cycle (i.e., 24 hours as the length of market horizon), in which all SGs are assumed to submit true offers, so the bilevel model reduces to a single-level problem as \eqref{eq:LL_model}. This will serve as a model to compute honest SCC bids for various SGs and identify critical buses that need to follow SCC constraints. 

First, we remove SCC-related terms from the model to simulate an energy-only market, i.e., with no guarantees on adequate SCC levels. From the solution of this problem, the resulting minimum SCC for each bus is obtained. Any SCC lower than $\textrm{I}_{b_{\textrm{lim}}}$ means that the corresponding bus is at risk and should in fact incorporate an explicit SCC constraint. After adding these SCC constraints, using a value of $\textrm{I}_{b_{\textrm{lim}}}=\textrm{5}~\textrm{p.u.}$ in all system buses as in \cite{chu2021short}, truthful bids for this service are calculated through the marginal unit pricing introduced in~\cite{chu2024pricing}, as explained in Section~\ref{sec:bilevel_model}. 

\begin{figure}[t]
    \centering
    \includegraphics[width=1\linewidth]{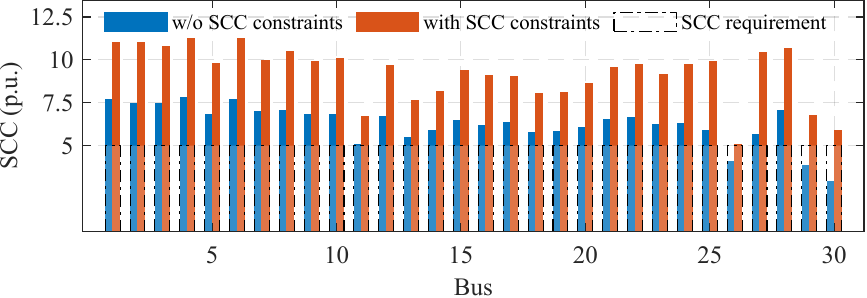}
    \caption{Minimum SCC levels in all buses with and without explicit constraints.}
    \label{fig:buses_low_SCC}
         %\vspace{-0.3cm}
\end{figure}

\begin{figure}[t]
    \centering
    \includegraphics[width=1\linewidth]{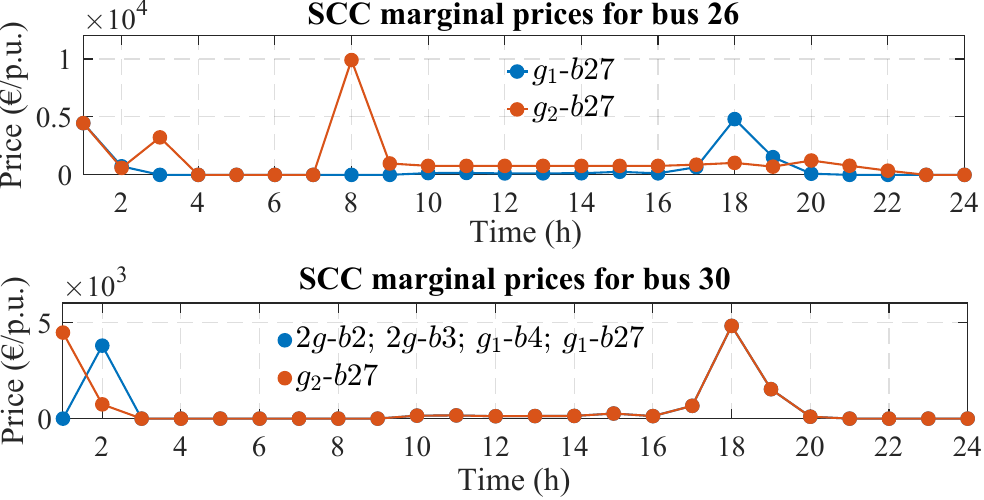}
    \caption{Marginal service prices for units providing SCC to critical buses in one-day operation. Prices in other cases are always zero, therefore not shown.}
    \label{fig:AS_offer_prices}
    % \vspace{-0.6cm}
\end{figure}

\begin{table}[t]
\centering
\caption{SCC Contributions from SGs to Critical Buses (p.u.)}
\setlength{\tabcolsep}{6pt}
{\fontsize{8pt}{15pt}\selectfont
\begin{tabular}{ccccc}
\toprule
\multirow{2}{*}{\vspace*{-3mm} SGs contributing SCC} & \multicolumn{3}{c}{ SCC captured by critical buses } \\
\cmidrule(lr){2-4}
& Bus 26 & Bus 29 & Bus 30 \\
\midrule   
$g_1$-$b$2/$g_2$-$b$2  & 2.75/2.75 & 2.95/2.95 & 2.91/2.92  \\
$g_1$-$b$3/$g_2$-$b$3  & 2.76/2.77 & 2.96/2.96 & 2.91/2.91  \\
$g_1$-$b$4/$g_2$-$b$4  & 2.77/2.77  & 2.97/2.99  & 2.91/2.91  \\
$g_1$-$b$5/$g_2$-$b$5  & 2.76/2.77     & 2.97/2.99     & 2.92/2.92         \\
$g_1$-$b$27/$g_2$-$b$27 & \cellcolor{gray!20}\textbf{3.32}/\cellcolor{gray!20}\textbf{3.34} & \cellcolor{gray!20}\textbf{3.63}/\cellcolor{gray!20}\textbf{3.67} & \cellcolor{gray!20}\textbf{3.65}/\cellcolor{gray!20}\textbf{3.86}  \\
%$g_1$-$b$30/$g_2$-$b$30 & 2.93/2.96 & 4.01/4.18 & 45.44  \\
\bottomrule
%\vspace{-0.5cm}
\end{tabular}
}
\label{table:SCC_contributions_to_buses}
\end{table}

The SCC distribution without explicit constraints is illustrated in Fig.~\ref{fig:buses_low_SCC}, confirming minimum SCC in buses \{26, 29, 30\} to be below the acceptable threshold. In this market setting, only units on buses \{2, 3\} are always committed to supply energy due to their low marginal costs, providing SCC meanwhile. It is worth noting that the SCC of bus 1 is beyond $\textrm{I}_{b_{\textrm{lim}}}$, though its local IBR can only provide up to 1 p.u. of SCC. That is because the bus is electrically close to online SGs, therefore benefiting from a safe level of SCC directly as a by-product of the energy market. However, the SCC on buses \{26, 29, 30\} may be insufficient within the day as they are at the edge of the system, being away from the typically committed thermal units (surrounding SGs are not always dispatched due to their high marginal costs).

The marginal SCC prices payable to different units for critical buses \{26, 29, 30\} are depicted in Fig.~\ref{fig:AS_offer_prices}, in which the prices for bus 30 are in fact the same for several units (blue line in the lower plot) due to their very similar SCC contribution to this bus. For the case of bus 26 (upper plot), only SGs in bus 27 would capture prices for SCC provision, whereas other SGs are not chosen to provide this service. This is due to the fact that 2$g$-$b$27 are closer to bus 26, providing a higher level of SCC which is already sufficient to secure that bus (as seen in Table~\ref{table:SCC_contributions_to_buses}), while other units provide a lower volume, resulting in those SCC contributions being less cost-effective. This explains that the elimination of those excess contributions does not incur higher costs. Vice versa, without SCC provision from $g_1$-$b$27 or $g_2$-$b$27, another local unit or more distant units with limited SCC contribution would have to be committed for meeting the security requirement, rendering higher costs which lead to these marginal SCC prices for units in bus 27. 

For bus 30 (lower plot), more than half of SGs participate in this service market capturing a certain price, which is generally lower than that for the case of bus 26 since most units provide low SCC due to long electrical distances, implying that the service from any of them can be replaced by any other. The marginal service price for bus 29 is zero, meaning that it shows sufficient SCC once other buses are secured, so no extra units would be required.

In summary, from this perfectly competitive market case it can be observed that only units $g_1$-$b$27 and $g_2$-$b$27 provide SCC to bus 26 at non-zero prices, while the market of bus 30 has multiple units at different electrical locations providing the service. Consequently, given the test system and operation data adopted, thermal units that do not offer SCC services to critical buses are naturally not profitable in the SCC market. The revenue term `$\lambda_{b,t}^{\textrm{SCC}}\textrm{k}_{bg}u_{g,t}$' is thus eliminated from their objective function in following studies.

\subsection{System Operation under Imperfect Competition}\label{System Operation in the Imperfectly Competitive Market}
The bilevel model introduced in Section~\ref{sec:bilevel_model} allows to represent the case where a power company may behave strategically in the market to increase profits. We assume this company to operate several SGs, so that  market power issues may be observed through analyzing different cases.

\begin{table}[t]
\centering
\caption{Evaluation of the Impact of Penalty Factor on Strategic Cases (Units of `$r^{\textrm{DG}}$' and `Profits' are \% and k\texteuro/day, respectively)}
{\fontsize{8pt}{12pt}\selectfont
\setlength{\tabcolsep}{2.2pt}
\begin{tabular}{c
                cc
                cc
                cc
                cc
                cc
                cc}
\toprule
& \multicolumn{10}{c}{Bus of strategic units} \\
\cmidrule(lr){3-12}
\multirow{2}{*}{\centering W} 
& \multirow{2}{*}{CPU} 
& \multicolumn{2}{c}{Bus 2} 
& \multicolumn{2}{c}{Bus 3} 
& \multicolumn{2}{c}{Bus 4} 
& \multicolumn{2}{c}{Bus 5} 
& \multicolumn{2}{c}{Bus 27} \\
\cmidrule(lr){3-4} \cmidrule(lr){5-6} \cmidrule(lr){7-8} \cmidrule(lr){9-10} \cmidrule(lr){11-12}
& Time (s)
& \(r^{\textrm{DG}}\) & Pro 
& \(r^{\textrm{DG}}\) & Pro 
& \(r^{\textrm{DG}}\) & Pro 
& \(r^{\textrm{DG}}\) & Pro 
& \(r^{\textrm{DG}}\) & Pro \\
\midrule
1    & 17.78   & 3.60  & 423  & 4.07  & 341  & 6.79  & 155  & 7.24  & 146  & 7.39  & 292  \\
\rowcolor{gray!20}
10   & 11.26    & -0.35 & 269  & 0.95  & 226  & 3.65  & 28   & 4.17  & 35   & 2.88  & 124  \\
100  & 10.14    & -0.38 & 262  & 0.87  & 208  & 3.59  & 13   & 4.10  & -3   & 2.78  & 103  \\
1000 & 15.83   & -0.38 & 261  & 0.87  & 197  & 3.59  & 12   & 4.10  & -4   & 2.78  & 103  \\
\bottomrule
\end{tabular}
}
\label{table:penalty_evaluation}
%\vspace{-0.2cm}
\end{table}

\begin{table}[t]
\centering
\caption{Averaged Strategic Multipliers of SGs in the Energy Market}
{\fontsize{8pt}{12pt}\selectfont
\setlength{\tabcolsep}{1.7pt} % 适当增大间距，避免拥挤
\begin{tabular}{c c c c c c c c c c c c}
\toprule
\multirow{2}{*}{Strategic units} 
& \multicolumn{2}{c}{Bus 2} 
& \multicolumn{2}{c}{Bus 3} 
& \multicolumn{2}{c}{Bus 4} 
& \multicolumn{2}{c}{Bus 5} 
& \multicolumn{2}{c}{Bus 27} \\
\cmidrule(lr){2-3} \cmidrule(lr){4-5} \cmidrule(lr){6-7} \cmidrule(lr){8-9} \cmidrule(lr){10-11}
& \(\hat{g}_1\) & \(\hat{g}_2\)
& \(\hat{g}_1\) & \(\hat{g}_2\)
& \(\hat{g}_1\) & \(\hat{g}_2\)
& \(\hat{g}_1\) & \(\hat{g}_2\)
& \(\hat{g}_1\) & \(\hat{g}_2\) \\
\midrule
\(\sum_{t}\beta_{\hat{g},t}^\textrm{m}/T\) 
& \cellcolor{gray!20}\textbf{1.71} & \cellcolor{gray!20}\textbf{1.67} & \cellcolor{gray!20}\textbf{1.54} & \cellcolor{gray!20}\textbf{1.29} & 1.05 & 1.03 & 1.00 & 1.00 & 1.04 & 1.00 \\
\bottomrule
\end{tabular}
}
\label{table:energy_market_power}
%\vspace{-0.4cm}
\end{table}

\begin{figure}[t]
    \centering
    \includegraphics[width=1\linewidth]{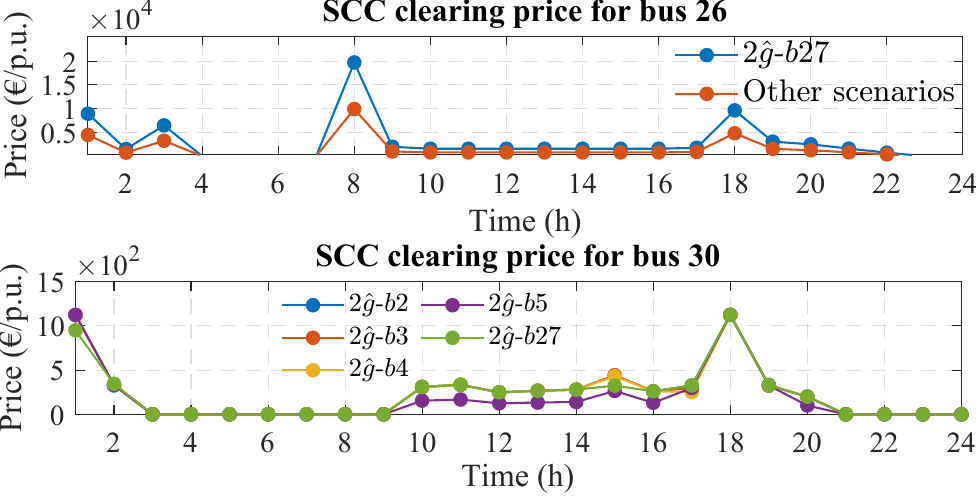}
    \caption{Shadow prices of SCC constraints for bus 26 and bus 30 when two strategic SGs are placed in a single bus. The price for bus 29 is zero, so not graphically shown.}
    \label{fig:SCC_prices}
        % \vspace{-0.3cm}
\end{figure}

\begin{table}[t]
\centering
\caption{Revenues of SGs for Providing SCC and Corresponding Service Payment (k\texteuro/day)}
\setlength{\tabcolsep}{3pt}
{\fontsize{7.9pt}{12pt}\selectfont
\begin{tabular}{ccccccc}
\toprule
\multirow{2}{*}{\vspace*{-3mm}\shortstack{Synchronous generators\\capturing shown revenues}} & \multicolumn{5}{c}{Location of strategic SGs (2$\hat{g}$-$bX$)} \\
\cmidrule(lr){2-6}
& Bus 2 & Bus 3 & Bus 4 & Bus 5 & Bus 27 \\
\midrule
2$g$-$b$2  & \textbf{11.13} & 11.08 & 10.96 & \cellcolor{gray!20}8.84 & 10.58 \\
2$g$-$b$3  & 11.12 & \textbf{11.08} & 10.96 & \cellcolor{gray!20}8.84 & 10.58 \\
2$g$-$b$4  & 5.56  & 5.54  & \textbf{10.97} & \cellcolor{gray!20}2.92 & 6.58 \\
2$g$-$b$5  & 0     & 0     & 0     & \cellcolor{gray!20}\textbf{0} & 0    \\
2$g$-$b$27 & 45.53 & 45.50 & 45.44 & \cellcolor{gray!20}44.38 & \cellcolor{red!20}\textbf{147.86} \\
\midrule
Payment for SCC services & 73.4  & 73.2  & 78.3   & \cellcolor{gray!20}65.0   & \cellcolor{red!20}\textbf{175.6}  \\
\bottomrule
%\vspace{-0.5cm}
\end{tabular}
}
\label{table:rev_AS_SGs}
\end{table}

\begin{figure}[t]
    \centering
    \includegraphics[width=1\linewidth]{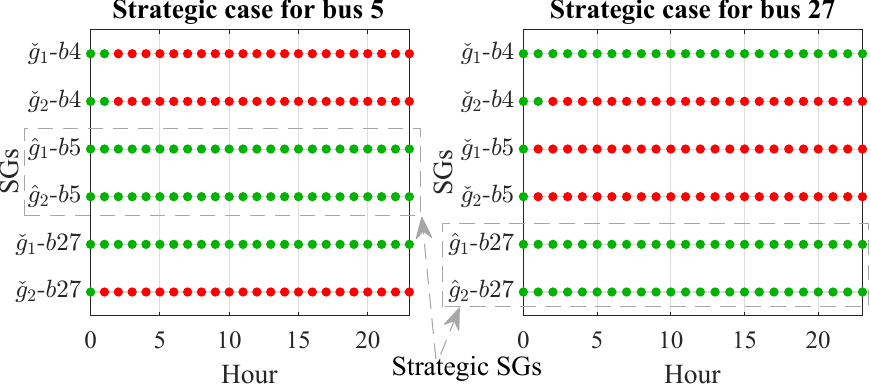}
    \caption{UC status of relevant SGs in cases `2$\hat{g}$-$b$5' (left) and `2$\hat{g}$-$b$27' (right). In these two cases, 2$\check{g}$-$b$2, 2$\check{g}$-$b$3 are always online, while 2$\check{g}$-$b$30 keep offline. \textcolor[rgb]{0,0.7,0}{$\bullet$}: Online; \textcolor{red}{$\bullet$}: Offline.}
    \label{fig:UC_24h}
        % \vspace{-0.3cm}
\end{figure}

\subsubsection{Two Strategic SGs in a Single Bus}\label{Two Strategic SGs in Single bus}
The strategic company in this setting owns two SGs placed in a single bus, thus producing five scenarios, i.e., buses \{2, 3, 4, 5, 27\} (bus 30 is not considered, as the units placed there are too expensive to be dispatched to provide services). The $r^{\textrm{DG}}$ \eqref{eq:ratio_DG} and profits (accounting for both energy and SCC markets) of the strategic company in all five cases are given in Table \ref{table:penalty_evaluation}, with values for the penalty term W of \{1, 10, 100, 1000\}. As the penalty factor increases, $r^{\textrm{DG}}$ and the profits decrease, whereas the computational time exhibits no significant variation, suggesting that the bilevel model can be solved with stable computational cost. It is noteworthy that $r^{\textrm{DG}}$ could be negative, as the primal LL problem is relaxed to be linear, thus having a bound potentially lower than the dual \cite{najman2021linearization}. From these results, we set going forward that the strategic firm gives equal weight to the DG (i.e., $\textrm{W}=\textrm{10}$) in all scenarios, since a noticeably small $r^{\textrm{DG}}$ can be achieved while yielding a substantial profit, the ultimate objective of the strategic company. It can also ensure that the weight between objectives in both levels are the same in all scenarios. Although higher profits could be obtained with the setting $\textrm{W}=\textrm{1}$, it actually suggests that the strategic firm cannot make a reasonable evaluation regarding its profitability \cite{ye2019incorporating}. While a larger W can only slightly reduce the gap, it noticeably lowers the profit at the upper level, thereby violating the criterion for selecting W. A comparable parameter selection strategy for balancing the objectives in \eqref{eq:pri_dual_obj} can be found in \cite{matamala2025strategic}, where the authors achieved similar performance in solving the bilevel problem, guaranteeing desirable profits for strategic agents with acceptable market clearing accuracy.

It should be noted that the value of W should be adjusted adaptively for different test systems to achieve a balanced trade-off among the aforementioned objectives.

While market power from SCC services is the focus of this work, we still briefly analyze market power in the energy clearing, by checking averaged bidding decisions, i.e., $\sum_{t}\beta_{\hat{g},t}^\textrm{m}/T$, shown in Table \ref{table:energy_market_power}. Units at buses \{2, 3\} are seen to strategically bid to maximize profits, whereas other expensive units normally offer true energy bids even if they are capable of misreporting them (as illustrated by the shaded area in comparison with other strategic scenarios; note that a bidding multiplier of 1 indicates a truthful bid). This is because there are significantly cheaper units with important capacity in the energy market, making high-cost units difficult to be dispatched. Consequently, the optimal solution for them is bidding honestly in order to clear the market. Market power in SCC services is analyzed next.
 
\textbf{\textit{Analysis of SCC prices:}} The SCC clearing prices for critical buses \{26, 30\} under five scenarios are illustrated in Fig.~\ref{fig:SCC_prices}, where the prices for bus 26 are significantly higher than those for bus 30. This is due to the differences in SCC marginal prices taken from Fig.~\ref{fig:AS_offer_prices} (used as the truthful bids in the bilevel model), and the fact that the SCC market of bus 26 has been monopolized by 2$\hat{g}$-$b$27, while other SGs are only able to provide SCC services to bus 30, as discussed in Section \ref{System Operation in the Perfectly Competitive Market}.

Specifically, for bus 26 the highest SCC price is achieved when 2$\hat{g}$-$b$27 act strategically (blue line). On the other hand, in strategic scenarios for buses \{2, 3, 4, 5\} (orange line), where units in bus 27 are non-strategic at the same time, the service price is still determined by the truthful bids from 2$\check{g}$-$b$27. This is due to the lack of substitutes in the market, i.e., 2$g$-$b$27 are the most effective units for providing SCC services here, therefore they are always chosen as the sole providers. In contrast, since several thermal units with similar  SCC contributions share the market for bus 30, implying that this market is relatively competitive rather than completely manipulated by certain resources, the SCC price varies according to which player is strategic. 

These results demonstrate that market power may be enhanced through occupying advantageous electrical positions, as these allow to provide SCC exclusively for critical buses, therefore achieving a greater manipulation of service prices.

\textbf{\textit{Payment and revenues for SCC services:}} The revenues of SGs in various strategic scenarios are specified in Table \ref{table:rev_AS_SGs}. This table shows revenues for generic SGs (first column) under different strategic SG pairs (right columns), with bold diagonal elements representing strategic SGs' revenues, and other values indicating non-strategic SGs' income.

The service fee that consumers would pay is shown on the last row of Table \ref{table:rev_AS_SGs}, which is equal to the sum of service revenues of all SGs. Since 2$\hat{g}$-$b$5 do not participate in the SCC market (as discussed in Section \ref{System Operation in the Perfectly Competitive Market}), being only strategic in the energy market, this is in fact the case where all SGs provide SCC services faithfully, leading to the lowest service revenues for SGs and consequently lowest payment by consumers (column in gray). When units at buses \{2, 3, 4\} are strategic, the fee varies slightly. However, for the strategic case `2$\hat{g}$-$b$27', total payments increase significantly due to the much higher revenues of 2$\hat{g}$-$b$27.

Except for 2$\hat{g}$-$b$27, the only SGs with a significant SCC contribution to bus 26, other strategic units do not manage to noticeably increase service revenues, as they can only profit in the market of bus 30. As explained earlier, bus 30 shows a competitive SCC market with clearing prices under different scenarios reaching similar values, as captured in the lower plot on Fig.~\ref{fig:SCC_prices}. On the other hand, the units in 2$\hat{g}$-$b$27 can strategically behave in the market of bus 26 in order to significantly increase their income, which could soar from 44.38 k\texteuro~(case `2$\hat{g}$-$b$5', when 2$g$-$b$27 are non-strategic) to 147.86 k\texteuro, more than tripled. 

It is interesting to notice that units 2$\hat{g}$-$b$27 manage to triple their revenues although the SCC price for bus 26 is only doubled in comparison to their non-strategic case, as seen in the upper plot on Fig.~\ref{fig:SCC_prices}. The reason for this further increase in income, not only related to price manipulation, is another detected strategic behavior by these units: they stay online for longer periods. %, offering a level of SCC to bus 26 which is higher than the grid-code requirement, thus further gaining more revenues. 
The hours when 2$g$-$b$27 are online for both cases `2$\hat{g}$-$b$5' and `2$\hat{g}$-$b$27' are depicted in Fig.~\ref{fig:UC_24h}. It can be observed that, in case `2$\hat{g}$-$b$5' (upper plot), $\check{g}_1$-$b$27 is constantly in operation to provide both energy and SCC services due to its slightly lower generation costs compared to $\check{g}_2$-$b$27, which is only committed in the first hour. This implies that the SCC requirements on critical buses can in fact be cost-effectively satisfied by just one thermal unit in bus 27, and other units such as 2$g$-$b$5. However, once $\hat{g}_2$-$b$27 behaves strategically (case `2$\hat{g}$-$b$27', lower plot), it would remain online for the whole market horizon, meanwhile changing the commitment of other SGs according to their SCC provision.

In summary, to increase revenues from SCC markets, certain strategic SGs could both manipulate service prices and commit for longer periods, further exacerbating the market distortion. This has been shown to be particularly relevant for the critical location in terms of SCC in this test network, bus 26, where the SCC market is monopolized by the electrically-closest generators placed in bus 27.

\subsubsection{Two Strategic SGs in Different Buses}\label{A Strategic Player Managing SGs in Electrically Distanced buses}

\begin{figure}[t]
    \centering
    \includegraphics[width=1\linewidth]{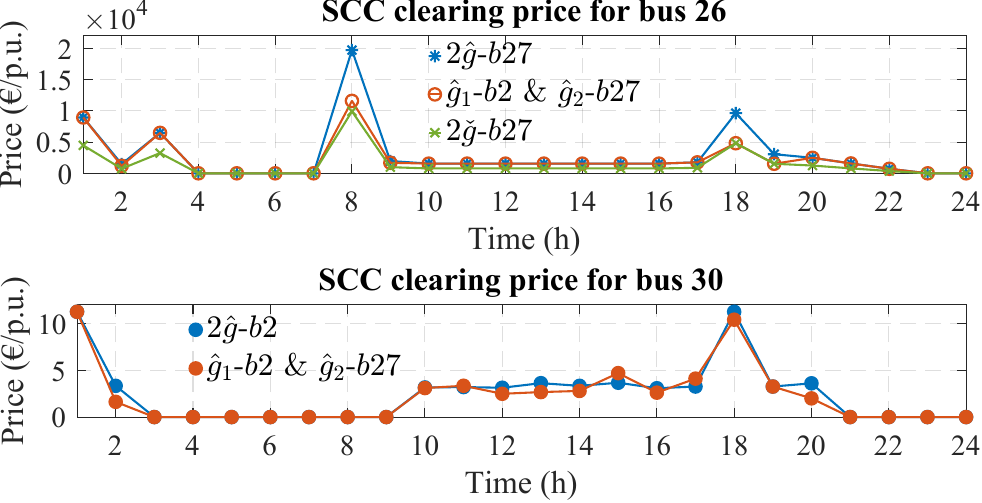}
    \caption{Comparison of SCC clearing prices for bus 26 and bus 30 when two strategic SGs are placed at different buses.}
    \label{fig:SCC_prices_distant}
                 %\vspace{-0.5cm}
\end{figure}

We now analyze the impact of a strategic company owning two generators placed in different system buses, instead of co-locating them in the same bus. Given that there exist many combinations of strategic SGs placed at different locations,  we properly select only two thermal units to carry out case studies. In this subsection, the scenario of strategic `$\hat{g}_1$-$b$2 \& $\hat{g}_2$-$b$27' is considered, as it represents general principles that other scenarios also follow. The penalty factor W is still set to 10, with $r^{\textrm{DG}}=\textrm{1.02\%}$.

\textbf{\textit{Analysis of SCC prices:}} The SCC prices for bus 26 and 30 under different scenarios are summarized in Fig.~\ref{fig:SCC_prices_distant}. For the case of bus 26 (upper plot), it can be seen that as the number of strategic SGs in bus 27 is progressively  reduced from two to zero, i.e., \{2$\hat{g}$-$b$27, blue line\}$\rightarrow$\{$\hat{g}_1$-$b$2 \& $\hat{g}_2$-$b$27, orange line\}$\rightarrow$\{2$\check{g}$-$b$27, green line\}, the price of SCC gradually decreases, as expected given the results in previous case studies. The strategic unit $\hat{g}_1$-$b$2 which replaces the first strategic generator in bus 27 is significantly less effective to increase prices for SCC in the market for bus 26. A similar trend is observable in the market for bus 30 (lower plot). 

These results show that it is best for the strategic company to concentrate their generators around the most critical location in terms of SCC requirements, which in this test network is bus 26. If the company were to make an investment in a second SG, assuming they already own one placed in bus 27 (the electrically closest generator bus to the critical bus 26), it shows to be a better decision to build it in the same bus 27 than splitting their generation assets geographically: indeed, case `$\hat{g}_1$-$b$2 \& $\hat{g}_2$-$b$27' reduces the company's capability to manipulate SCC prices.

\textbf{\textit{Payment and revenues for SCC services:}} Accordingly, the SCC service payment chargeable to consumers would change as follows: \{175.6 k\texteuro$\mid$2$\hat{g}$-$b$27\}$\rightarrow$\{96.2 k\texteuro$\mid$$\hat{g}_1$-$b$2 \& $\hat{g}_2$-$b$27\}, showing a significant reduction. As for the SCC revenues of relevant SGs, non-strategic $\check{g}_1$-$b$27 only earns 10.1 k\texteuro, while $\hat{g}_2$-$b$27 gets 59.1 k\texteuro, with the total revenues being 69.2 k\texteuro, noticeably lower than 147.86 k\texteuro~in the case `2$\hat{g}$-$b$27' (which was shown in Table \ref{table:rev_AS_SGs}). 

In summary, market power issues exacerbate when the number of strategic units in the same critical location increases, rather than a company owning generators in different buses. This suggests that procuring SCC by other means, e.g., by investing in a synchronous compensator owned by the Transmission System Operator (TSO) in that critical location, could significantly mitigate market power. Then, the SCC market may be maintained for the rest of the system, as it shows to be fairly competitive in the test network considered.

\subsection{Potential Countermeasures to Mitigate Market Power }\label{Potential Countermeasures to Mitigate Market Power}

Countermeasures to limit market power in SCC services can be broadly categorized into two types: market and technology-based.

\subsubsection{ Market-Based Measures }\label{Market-Based Measures}

\textbf{\textit{ Upper limit of secure SCC:}} An upper bound on SCC constraints can be imposed at critical buses to enable greater participation of units from different locations in the market, thereby maintaining SCC within a limited range. For example, setting an appropriate SCC limit at bus 26 could enforce SGs at bus 27, which contribute significantly to SCC, to be shut down, while other units with smaller SCC contributions are dispatched as substitutes. However, determining an appropriate upper bound remains challenging. If the limit is set too low, essential SGs may be forced offline, especially given that SCC is provided in a discrete manner. Further analysis on this approach is necessary to conclude its suitability for a given system.

\textbf{\textit{ Service price cap:}} Setting a cap on SCC service prices could directly limit market power. However, determining an appropriate cap would require accurate evaluation of service offers submitted by market participants to ensure efficient market clearing. In this study, the bidding multiplier is capped at 2, which is adopted as a conservative upper-bound assumption to fully characterize the maximum potential of strategic SGs to exercise market power. A key issue is that, in practice, the bidding decision does not necessarily reflect the true marginal cost perceived by the service provider. Even with a relatively low bid, an agent may claim a higher marginal service cost; conversely, a higher bid may be associated with a lower actual cost. Ultimately, system operators can only observe the submitted service offers.
Therefore, a critical challenge is that some power companies may somehow misreport their offers, which makes it difficult for regulators or system operators to achieve cost-effective system operation \cite{rebours2007survey}.

\textbf{\textit{ Demand response:}} This market mechanism allows consumers or load aggregators to flexibly curtail or shift their loads when SCC service prices rise or when SGs with market power operate for extended periods. Such flexible load adjustments not only have the potential to restrain the strategic commitment behavior of these SGs (since lower loading conditions may cause certain units to go offline), but also enable consumers to receive economic compensation for their load regulation efforts, thereby directly alleviating the adverse economic impacts of strategic SGs' behavior on the demand side. A preliminary analysis on the role of demand response in SCC-constrained UC has been conducted in \cite{wang2025analyzing}, laying the foundation for further investigating its impact on the market power of SCC services.

\subsubsection{Technology-Based Measures }
\begin{table}[t]
\centering
\caption{Revenues of SGs for Providing SCC and Corresponding Service Payment when Synchronous Condensers are Installed on Bus 26 (k\texteuro/day)}
\setlength{\tabcolsep}{3pt}
{\fontsize{7.9pt}{12pt}\selectfont
\begin{tabular}{ccccc}
\toprule
\multirow{2}{*}{\vspace*{-3mm}\shortstack{Synchronous generators\\capturing shown revenues}} & \multicolumn{4}{c}{Location of strategic SGs (2$\hat{g}$-$bX$)} \\
\cmidrule(lr){2-5}
& Bus 2 & Bus 3 & Bus 4 & Bus 5  \\
\midrule
2$g$-$b$2  & 11.13 & 11.08 & 10.96 & 8.84  \\
2$g$-$b$3  & 11.12 & 11.08 & 10.96 & 8.84  \\
2$g$-$b$4  & \textbf{6.32}  & \textbf{6.13}  & 10.97 & \textbf{3.25} \\
2$g$-$b$5  & 0     & 0     & 0     & 0     \\
2$g$-$b$27 & \textbf{5.47} & \textbf{5.19} & \textbf{5.06} & \textbf{4.53}  \\
\midrule
Payment for SCC services & \textbf{34.04}  & \textbf{33.48}  & \textbf{37.95}   & \textbf{25.46}   \\
\bottomrule
%\vspace{-0.5cm}
\end{tabular}
}
\label{table:SCC revenue support devices}
\end{table}

\textbf{\textit{SCC support devices:}} A feasible solution is to deploy local SCC response devices or other auxiliary technologies, such as synchronous condensers. 

Table~\ref{table:SCC revenue support devices} presents the SCC revenues of generators and the ancillary service costs borne by consumers after synchronous condensers that provide the same level of SCC as 2$g$-$b$27 are installed. It is evident that if a synchronous condenser is installed on bus 26, the market power analyzed in this paper will be mitigated significantly (compared with Table~\ref{table:rev_AS_SGs}, the SCC revenues of 2$g$-$b$27 and consumer payments decrease substantially), as the SCC level for that bus could be satisfied without requiring further SCC services. Generators at buses 2 and 3 remain online primarily to meet energy demand, and thus their SCC revenues stay unchanged. By contrast, 2$g$-$b$4 achieve slightly higher SCC revenues relative to Table~\ref{table:rev_AS_SGs}. This is attributed to the high operating costs of 2$g$-$b$27, which no longer supply critical SCC support to bus 26. Consequently, lower‑cost generators at bus 4 are dispatched more frequently to satisfy energy demand, enabling them to obtain greater SCC revenues.  

However, it is then relevant to analyze who should own and be responsible for the investment cost in these synchronous condensers.

\textbf{\textit{System topology:}} Analyzing the electrical topology of the system to assess short-circuit risks at each bus and to identify locations that may give rise to market power could also be an effective approach. In this way, investment decisions can better balance system benefits against the number of units installed at such buses, and the network topology could also be reconfigured to mitigate market power. The latter approach would entail additional complexity for system operators, which should be further analyzed.

\subsection{Potential Impact of Model Settings on Market Power in SCC Services }\label{Potential Impact of System Operation on Market Power in SCC Services}
In this subsection, we clarify the potential impact of system assumptions in this paper on the manifestation of market power for SCC. 

\subsubsection{Uniform SCC Threshold} 
\begin{table}[!t]
\centering
\caption{SCC Revenue Variation to SCC Security Thresholds}
\setlength{\tabcolsep}{2.5pt}
{\fontsize{8pt}{12pt}\selectfont
\begin{tabular}{lcccc}
\toprule
\qquad \qquad $\textrm{I}_{26_{\textrm{lim}}}$ (p.u.)  & 4.40 & 4.70 & 5.00 & 5.30  \\
\midrule
SCC revenues of 2$\hat{g}$-$b$27 (k\texteuro/day)  & \cellcolor{gray!20}112.37 & 147.86 & 147.86 & 147.86 \\
SCC revenues of 2$\check{g}$-$b$27 (k\texteuro/day)  & \cellcolor{gray!20}36.52 & 44.38 & 44.38 & 44.38 \\
\bottomrule
\end{tabular}
}
\label{table:Market Power Variation to SCC Security Thresholds}
%\vspace{-0.3cm}
\end{table}
In the present work, a uniform secure SCC threshold, $\textrm{I}_{b_{\textrm{lim}}}$, is adopted for all buses. In practice, however, the minimum acceptable SCC level may vary depending on the bus type due to factors such as protection configurations, and it should be confirmed before the system scheduling. This work is not intended to determine $\textrm{I}_{b_{\textrm{lim}}}$ for each individual bus, but to propose a market framework that enables relevant agents to examine market power using given SCC thresholds. In order to provide a practical and conservative SCC security, $\textrm{I}_{b_{\textrm{lim}}} = 5~\textrm{p.u.}$ is adopted in the case studies, which guarantees strong grid conditions at the buses of interest, as specified in \cite{chu2021short}.

In this case, a sensitivity analysis of SCC requirements at bus 26 to capture variations in market power is still worthy of investigation. Table~\ref{table:Market Power Variation to SCC Security Thresholds} presents the SCC revenues of 2$g$-$b$27 under different SCC security levels. It can be observed that within the range from 4.70 p.u. to 5.30 p.u., the SCC revenues of these generators remain unchanged, as no other generators in the market can substitute their provided SCC capacity. However, when the threshold requirement drops to 4.40 p.u., alternative generators become available to satisfy security constraints (Table~\ref{table:SCC_contributions_to_buses} shows that the SCC capacity supplied by other generators to bus 26 can cover a capacity gap of 0.60 p.u.). Consequently, the SCC revenues of 2$\hat{g}$-$b$27 decrease when they exercise market power. Nevertheless, their profits are still three times higher compared with their non‑strategic operation scenario, i.e., 2$\check{g}$-$b$27, which shows the persistence of substantial market power.

This analysis demonstrates that the uniform SCC threshold assumption does not undermine the results of this study, as the local or remote coupling between SGs and buses (the fundamental driver of market power) is inherently dependent on the electrical distance, rather than on $\textrm{I}_{b_{\textrm{lim}}}$. It is worth noting, however, that this setting may instead lead to an underestimation of market power for SGs located near buses with stricter practical SCC requirements (e.g., grid-following IBR buses that may require a higher level of SCC). Such buses rely more on electrically proximate SGs, resulting in lower service substitutability. In the system studied in this paper, these factors would amplify the market power of SGs at bus 27 with respect to bus 26, underscoring the importance of carefully designing the SCC ancillary service market.

\subsubsection{Approximated SCC Representation}
Another point worth noting is that the approximated SCC levels \eqref{eq:define_SCC_constraints_linearized} may introduce deviations in market outcomes, such as strategic behaviors and SCC clearing prices. Nevertheless, this bias remains within a controllable range and affects system-wide SCC levels rather than those at specific buses, implying that the core conclusion that market power is fundamentally driven by system topology and location of the different assets would not be altered.

From a system stability perspective, this approximation error is not of concern, as the condition $ \textrm{I}_{b_\textrm{lim}} \leq I_{b_\textrm{L}} \leq I_{b_\textrm{SC}}$ holds, indicating that the classification inherent in the approximation process introduces an implicit safety margin into the system. This observation is further supported by the training performance of the Type-II error computed by \eqref{eq:error} and reported in the last column of Table~\ref{table:SCC_validation}, where the mean value of this error remains negative.

\subsubsection{Uncertainty of Wind Generation }\label{Uncertainty of Wind Generation}
\begin{table}[!t]
\centering
\caption{SCC Revenue Variation to Wind Power Resources}
\setlength{\tabcolsep}{1.4pt}
{\fontsize{8pt}{12pt}\selectfont
\begin{tabular}{lccccc}
\toprule
 Wind power capacity (MW)  & 550 & 650 & 750 & 850 & 950  \\
\midrule
SCC revenues of 2$\hat{g}$-$b$27 (k\texteuro/day)  & 147.86 & 147.86 & 147.86 & \cellcolor{gray!20}97.54 & \cellcolor{gray!20}71.93 \\
SCC revenues of 2$\check{g}$-$b$27 (k\texteuro/day)  & 44.38 & 44.38 & 44.38 & \cellcolor{gray!20}38.87 & \cellcolor{gray!20}25.26 \\
\bottomrule
\end{tabular}
}
\label{table:SCC Revenue Variation to Wind Power Resources}
%\vspace{-0.3cm}
\end{table}
The impact of wind power capacity on market power in SCC services primarily arises through its coupling with the commitment of SGs. This is because, on the one hand, wind-driven IBR contribute significantly less to SCC than SGs and would be therefore insufficient for themselves to exert market power; on the other hand, the net load (system energy demand net of available wind power) directly affects the commitment status of SGs, and consequently the supply of SCC. 

To verify this viewpoint, market power under different levels of available wind power resources is analyzed herein. Table~\ref{table:SCC Revenue Variation to Wind Power Resources} shows the SCC revenues of generators at bus 27 under different wind turbine capacities. It can be observed that IBR installed capacities ranging from 550 MW to 750 MW have no impact on the SCC revenues of these two generators. This is because the reduced wind power resources can be offset by their rated output (as shown in Table~\ref{table:SGs_para}), meaning they can still remain online and are able to exercise market power. Nevertheless, with higher wind power penetration (from 850 MW to 950 MW), the generators are forced offline during certain periods to maintain power balance. Consequently, they cannot sustain their market‑power exercise, leading to lower SCC revenues that vary inversely with growing wind power availability. However, exhibiting market power still yields substantial profits for them compared with non‑strategic operation.

A phenomenon one can expect is that, under high net-load conditions, SGs are committed more frequently to satisfy energy demand and may simultaneously exert market power in SCC. By contrast, when load demand is low or wind power availability is high, SGs with lower costs and appropriate SCC capacity are more likely to exert market power, as they are preferentially dispatched by the system operator, while other higher-cost units may not be dispatched at all.

Consequently, the uncertainty of wind power output affects market power in SCC services in an indirect manner. Specifically, as long as wind generation remains at a level that does not alter the commitment status of SGs, its uncertainty would not, in theory, significantly change market outcomes. However, when available wind resources are close to the level at which certain SGs need to switch their operating states, such uncertainty may substantially mitigate or exacerbate market power, depending on whether the SGs exerting market power are switched to offline or online.

\subsubsection{Single-Node Energy Price }
The UC model in this paper adopts the simplifying assumption of single-node electricity price. When congestion exists, the spatial heterogeneity of Locational Marginal Prices (LMPs) is coupled with the locational characteristics of SCC prices; however, this coupling would be implicit. This is because the SCC price is directly related to binary unit commitment variables, whereas congestion is capacity-related and only indirectly linked to unit commitment decisions. Future research should explicitly characterize this relationship.

\subsubsection{Single-Day System Scheduling }
For longer multi-day scheduling horizons with more diverse system operating conditions, actual generating units will optimize their cross-period schedules. This may either mitigate or enhance the short-term revenue gains of strategic agents, depending on the actual electricity demand and available wind power, as discussed in Sections \ref{Market-Based Measures} and \ref{Uncertainty of Wind Generation}, but cannot eliminate the inherent market power arising from the grid topological heterogeneity in SCC provision.

\section{Conclusion and Future Work}\label{Conclusion}
A bilevel model has been introduced to explore potential market power issues in SCC services, allowing to capture strategic behaviors of some agents. Given the non-convex nature of the lower-level problem, as it contains binary commitment variables for SGs, directly replacing it by its KKT conditions is not a suitable approach for eventually obtaining an equivalent single-level problem. Therefore, a primal-dual formulation is employed instead to solve the model, which has been shown to work effectively in this setting.

Case studies based on a modified IEEE 30-bus system have revealed how strategic units exert market power in SCC services to achieve higher revenues. Generators located next to critical buses in terms of SCC requirements have been shown to effectively manipulate prices for this service in their own benefit. Furthermore, they could also stay online for a longer time, in order to both capture the higher prices for a sustained period and create barriers for other providers of this service. Consequently, it is demonstrated that these generators could achieve three times the revenue compared with scenarios where they do not exert market power. This has the double effect of limiting the profits of non-strategic units and leading to a higher service cost to consumers, highlighting the need to carefully design SCC markets.

The core novelty of this paper lies in the rigorous primal‑dual reformulation of an SCC‑constrained bilevel model for quantifying market power in SCC ancillary service markets, with key practical implications summarized as follows. Detailed market power studies for specific power systems should be carried out before determining if a market for SCC is suitable for those systems, given that alternative SCC procurement mechanisms (e.g., cost-based contracts, grid codes) would also be feasible complements for bus-specific SCC needs. Given the issues identified with regard to imperfect competition in SCC markets, potential countermeasures could be considered at both the market and technology levels, such as limiting the clearing price of SCC services, weighing the negative impacts of installing new units in buses that are prone to trigger market power when planning the system. Investing in synchronous condensers at proper locations would also be effective, making it relevant to analyze who should own and bear the investment costs. 

The main directions for future work include the following. First, further validation of the aforementioned market‑power mitigation countermeasures would be conducted. Second, future research would incorporate market participants that influence SCC provision, such as load aggregators, given that these entities may engage in strategic behavior for self‑interest and potentially alter market power in SCC service markets. Beyond these two core strands, follow‑up studies would also investigate power systems of different scales under bus‑specific SCC security requirements, so as to identify critical SCC nodes in more complex power systems, as SCC market power is inherently topology-driven, and its quantitative magnitude may vary across specific networks.

\appendix
\section{Optimization for Approximation of SCC Constraints}
This procedure comprises an enumeration for system operating points and a minimization problem that tries to reduce the error led by the approximation. This optimization is operated as follows:
\begin{subequations}\label{eq:opt_problem}
\begin{align}
& \min_{\mathcal{K}} \quad \sum_{\omega \in \Omega_2}
\left( I_{b_\textrm{L}}^{(\omega)} - I_{b_\textrm{SC}}^{(\omega)} \right)^2
\label{eq:opt_problem_a} \\
& \text{subject to:} \nonumber \\
& I_{b_\textrm{L}}^{(\omega)} < \textrm{I}_{b_{\textrm{lim}}}
\quad \forall \, \omega \in \Omega_1
\label{eq:opt_problem_b} \\
& I_{b_\textrm{L}}^{(\omega)} \ge \textrm{I}_{b_{\textrm{lim}}}
\quad \forall \, \omega \in \Omega_3
\label{eq:opt_problem_c} \\
& \Omega = \Omega_1 \cup \Omega_2 \cup \Omega_3
\label{eq:opt_problem_d} \\
& \Omega_1 =
\left\{
\omega \in \Omega \,\middle|\,
I_{b_\textrm{SC}}^{(\omega)} < \textrm{I}_{b_{\textrm{lim}}}
\right\}
\label{eq:opt_problem_e} \\
& \Omega_2 =
\left\{
\omega \in \Omega \,\middle|\,
\textrm{I}_{b_{\textrm{lim}}}
\le I_{b_\textrm{SC}}^{(\omega)}
< \textrm{I}_{b_{\textrm{lim}}} + \nu
\right\}
\label{eq:opt_problem_f} \\
& \Omega_3 =
\left\{
\omega \in \Omega \,\middle|\,
\textrm{I}_{b_{\textrm{lim}}} + \nu
\le I_{b_\textrm{SC}}^{(\omega)}
\right\}
\label{eq:opt_problem_g}
\end{align}
\end{subequations}
where \eqref{eq:opt_problem_a} is the objective for a minimum training error. $\Omega$ stands for all enumerated system operating points. Eqs.~\eqref{eq:opt_problem_b} and \eqref{eq:opt_problem_e} enable the correct classification for all sub-limit SCC level. A positive parameter $\nu$ is introduced to define $\Omega_2$ and $\Omega_3$ (as specified in \eqref{eq:opt_problem_f}–\eqref{eq:opt_problem_g}), so that all data points in $\Omega_3$ are classified correctly and any misclassification is restricted to subset $\Omega_2$. Furthermore, $\nu$ should be set as small as possible while ensuring the feasibility, so that an approximation with desired training performance can be achieved, as analyzed in Section \ref{Validation of Short Circuit Current Constraints}.

This classification only serves to train the coefficients in \eqref{eq:SCC_constraints_linearlized} for fitting the actual SCC, meaning that it is in fact an offline process to the bilevel optimization model. An example for the training process used in this work can be accessed in \cite{Code}.

\section{Marginal Unit Pricing for Computing SCC Service Offers}
We compute the SCC service offer $\textrm{O}_{g,b,t}^\textrm{SCC}$ via a marginal unit pricing \cite{chu2024pricing}, from which the economic value of SCC from SG $g$ for bus $b$, i.e., $\textrm{k}_{bg}$, is obtained as the commitment price (this indicates that the service offer is implicitly expressed as $\textrm{O}_{g,b,t}^\textrm{SCC}/\textrm{k}_{bg}$, with the unit being \texteuro/p.u.). This computation implies obtaining the difference in objective values between two problems. First, problem \eqref{eq:LL_model} is solved as an energy-only UC without SCC-related terms in the objective, denoted as $f^*(\textbf{x})$. Second, the SCC provision $\textrm{k}_{bg}$ is eliminated, as illustrated by \eqref{eq:marginal_unit_pricing}, then calculating hourly operation costs of the modified dispatch problem, denoted as $f(\textbf{x})$. The difference in those two solutions (i.e., $ f(\textbf{x}) - f^*(\textbf{x})$) represents the cost of forcing other units to startup for compensating the removed SCC, i.e., the marginal service costs of SCC. This whole process is further explained through a numerical example in Section~\ref{System Operation in the Perfectly Competitive Market}. 
\begin{subequations}\label{eq:marginal_unit_pricing}
    \begin{align}
   & \displaystyle \min  \quad f(\textbf{x}) \label{eq:marginal_unit_obj} \\
    & \textrm{s.t.} ~ \sum_{g}\Delta \textrm{k}_{bg}u_{g,t} + \sum_{c}\textrm{k}_{bc}\upalpha_{c,t} \ge \textrm{I}_{b_{\textrm{lim}}} | \{  \Delta \textrm{k}_{bg} | \textrm{k}_{bg},0 \}\label{eq:pricing_cons_SCC}  \\    
   & \qquad g(\textbf{x})\leq 0 \label{eq:marginal_unit_ineq_con}  \\
   &  \qquad h(\textbf{x})=0 \label{eq:marginal_unit_eq_con}
    \end{align}
\end{subequations}
where $\textbf{x}$ represents decision variables. Eq.~\eqref{eq:pricing_cons_SCC} shows that the variable SCC coefficient $\Delta \textrm{k}_{bg}$ is removed by setting $\Delta \textrm{k}_{bg} = 0$. Eqs.~\eqref{eq:marginal_unit_ineq_con}-\eqref{eq:marginal_unit_eq_con} indicate all other inequality and equality constraints remain unchanged.

Furthermore, since the commitment of SGs has an impact in energy provision and SCC contribution, these two services are highly coupled. As a result, some SCC contributions come as by-products of power balance or securing certain buses rather than given ones, thereby being essentially free (i.e., $\textrm{O}_{g,b,t}^\textrm{SCC}$ is zero). In summary, the value of this two-step procedure for computing $\textrm{O}_{g,b,t}^\textrm{SCC}$ is to recognize which SCC contributions to critical buses come with non-zero service costs.

\section{McCormick Envelopes}
\label{app:mccormick}
The McCormick envelopes with appropriate bounds, as suggested in \cite{ye2019incorporating}, are constructed for the relevant bilinear terms, as presented below.

\textbf{\textit{Envelope 1}}: The term $\lambda^\textrm{E}_{t}  P_{\hat{g},t}$ that follows $0 \leq \lambda^\textrm{E}_{t} \leq \overline{\lambda}^{\textrm{E}}$ can be reformulated as:
\begin{subequations}\label{eq:McCormick_lambdaP}
\begin{alignat}{2}
& \lambda_t^{\textrm{E}}\textrm{P}_{\hat{g}}^{\textrm{min}} \leq  z^{{\textrm{Re}^\textrm{E}}}_{\hat{g},t} \leq \overline{\lambda}^{\textrm{E}}P_{\hat{g},t} +\textrm{P}_{\hat{g}}^{\textrm{min}}(\lambda_t^{\textrm{E}}-\overline{\lambda}^{\textrm{E}})  \label{eq:McCormick_lambdaP_P_definition} \\
& \overline{\lambda}^{\textrm{E}}P_{\hat{g},t} +\textrm{P}_{\hat{g}}^{\textrm{max}}(\lambda_t^{\textrm{E}}-\overline{\lambda}^{\textrm{E}}) \leq z^{{\textrm{Re}^\textrm{E}}}_{\hat{g},t} \leq \lambda_t^{\textrm{E}}\textrm{P}_{\hat{g}}^{\textrm{max}} \label{eq:McCormick_lambdaP_con_2}
\end{alignat}
\end{subequations}
where $z^{{\textrm{Re}^\textrm{E}}}_{\hat{g},t}=\lambda^\textrm{E}_{t} P_{\hat{g},t}$. The upper bound should consider marginal costs that may be reported by both strategic and non-strategic units. Therefore, $\overline{\lambda}^{\textrm{E}}=\mathrm{max}~\{\textrm{O}_{\hat{g}}^\textrm{m} \overline{\upbeta}^\textrm{m}_{\hat{g}}, \textrm{O}_{\check{g}}^\textrm{m} \mid \forall \hat{g}, \forall \check{g}\}$. 

\textbf{\textit{Envelope 2}}: For term $\lambda_{b,t}^{\textrm{SCC}}\textrm{k}_{b\hat{g}} u_{\hat{g},t}$ complying with constraint $0 \leq \lambda_{b,t}^{\textrm{SCC}} \leq \overline{\lambda}_{b,t}^{\textrm{SCC}}$, it is restated as:
\begin{subequations}\label{eq:big_M_lambdaF_binary}
\begin{align}
& 0 \leq z^{{\textrm{Re}^\textrm{SCC}}}_{\hat{g},b,t} \leq \overline{\lambda}_{b,t}^{\textrm{SCC}}\textrm{k}_{b\hat{g}} u_{\hat{g},t} \label{eq:big_M_lambdaF_binary_1}  \\
& \lambda_{b,t}^{\textrm{SCC}}\textrm{k}_{b\hat{g}}-\overline{\lambda}_{b,t}^{\textrm{SCC}}\textrm{k}_{b\hat{g}}(1-u_{\hat{g},t}) \leq z^{{\textrm{Re}^\textrm{SCC}}}_{\hat{g},b,t} \leq \lambda_{b,t}^{\textrm{SCC}}\textrm{k}_{b\hat{g}} \label{eq:big_M_lambdaF_binary_2} 
\end{align}
\end{subequations}
where $z^{{\textrm{Re}^\textrm{SCC}}}_{\hat{g},b,t}=\lambda_{b,t}^{\textrm{SCC}}\textrm{k}_{b\hat{g}} u_{\hat{g},t}$. Similarly, the upper limit is set as $\overline{\lambda}_{b,t}^{\textrm{SCC}}= \mathrm{max}~\{\overline{\upbeta}^\textrm{SCC}_{\hat{g},b}\textrm{O}_{\hat{g},b,t}^\textrm{SCC}/\textrm{k}_{b\hat{g}}, \textrm{O}_{\check{g},b,t}^\textrm{SCC}/\textrm{k}_{b\check{g}} \mid \forall \hat{g}, \forall \check{g} \}$.

\textbf{\textit{Envelope 3}}: The nonlinear term $\textrm{O}_{\hat{g},b,t}^\textrm{SCC}\beta_{\hat{g},b,t}^\textrm{SCC} u_{\hat{g},t}$, constrained by \eqref{eq:UL_cons_AS}, is remodeled as:
\begin{subequations}\label{eq:McCormick_OPu}
\begin{alignat}{2}
& z^{{\textrm{Bid}^\textrm{SCC}}}_{\hat{g},b,t} \ge \textrm{O}_{\hat{g},b,t}^\textrm{SCC} u_{\hat{g},t} \label{eq:McCormick_OPu_1} \\
& z^{{\textrm{Bid}^\textrm{SCC}}}_{\hat{g},b,t} \ge \textrm{O}_{\hat{g},b,t}^\textrm{SCC}(\overline{\upbeta}^\textrm{SCC}_{\hat{g},b}(u_{\hat{g},t}-1) +\beta_{\hat{g},b,t}^\textrm{SCC}) \label{eq:McCormick_OPu_2} \\
& z^{{\textrm{Bid}^\textrm{SCC}}}_{\hat{g},b,t} \leq \textrm{O}_{\hat{g},b,t}^\textrm{SCC} \overline{\upbeta}^\textrm{SCC}_{\hat{g},b}u_{\hat{g},t}  \label{eq:McCormick_OPu_3} \\
& z^{{\textrm{Bid}^\textrm{SCC}}}_{\hat{g},b,t} \leq \textrm{O}_{\hat{g},b,t}^\textrm{SCC}(u_{\hat{g},t}+ \beta_{\hat{g},b,t}^\textrm{SCC}-1) \label{eq:McCormick_OPu_4} 
\end{alignat}
\end{subequations}
where $z^{{\textrm{Bid}^\textrm{SCC}}}_{\hat{g},b,t}=\textrm{O}_{\hat{g},b,t}^\textrm{SCC}\beta_{\hat{g},b,t}^\textrm{SCC} u_{\hat{g},t}$.

\textbf{\textit{Envelope 4}}: Likewise, term $ \textrm{O}_{\hat{g}}^\textrm{m} \beta_{\hat{g},t}^\textrm{m} P_{\hat{g},t}$, following \eqref{eq:UL_cons_energy}, is derived as:
\begin{subequations}\label{eq:McCormick_OKP}
\begin{alignat}{2}
& z^{{\textrm{Bid}^\textrm{E}}}_{\hat{g},t} \ge \textrm{O}_{\hat{g}}^\textrm{m} (P_{\hat{g},t}+ \textrm{P}_{\hat{g}}^{\textrm{min}}(\beta_{\hat{g},t}^\textrm{m}-1) )\label{eq:McCormick_OKP_1} \\
& z^{{\textrm{Bid}^\textrm{E}}}_{\hat{g},t} \ge \textrm{O}_{\hat{g}}^\textrm{m} ( \overline{\upbeta}^\textrm{m}_{\hat{g}}P_{\hat{g},t} +\textrm{P}_{\hat{g}}^{\textrm{max}}(\beta_{\hat{g},t}^\textrm{m}-\overline{\upbeta}^\textrm{m}_{\hat{g}}))\label{eq:McCormick_OKP_2} \\
&  z^{{\textrm{Bid}^\textrm{E}}}_{\hat{g},t} \leq \textrm{O}_{\hat{g}}^\textrm{m} (\overline{\upbeta}^\textrm{m}_{\hat{g}}P_{\hat{g},t} +\textrm{P}_{\hat{g}}^{\textrm{min}}(\beta_{\hat{g},t}^\textrm{m}-\overline{\upbeta}^\textrm{m}_{\hat{g}}))\label{eq:McCormick_OKP_3} \\
& z^{{\textrm{Bid}^\textrm{E}}}_{\hat{g},t} \leq \textrm{O}_{\hat{g}}^\textrm{m} (P_{\hat{g},t}+ \textrm{P}_{\hat{g}}^{\textrm{max}}(\beta_{\hat{g},t}^\textrm{m}-1)) \label{eq:McCormick_OKP_4} 
\end{alignat}
\end{subequations}
where $z^{{\textrm{Bid}^\textrm{E}}}_{\hat{g},t}=\textrm{O}_{\hat{g}}^\textrm{m}\beta_{\hat{g},t}^\textrm{m} P_{\hat{g},t} $.

\begin{table}[!t]
\centering
\caption{Key Parameters for SCC Constraints and Bilevel Model}
\setlength{\tabcolsep}{2.5pt}
{\fontsize{8pt}{12pt}\selectfont
\begin{tabular}{lccccccc}
\toprule
Parameters & $S_B$ & $\textrm{I}_c$ & $\nu$ & $\textrm{I}_{b_{\textrm{lim}}}$ & $\overline{\upbeta}^\textrm{m}_{\hat{g}}$ & $\overline{\upbeta}^\textrm{SCC}_{\hat{g},b}$ & W \\
\midrule
Values & 100 MVA & 1.00 p.u. & 0.1 p.u. & 5.00 p.u. & 2 & 2 & 10 \\
\bottomrule
\end{tabular}
}
\label{table:Key Parameters}
%\vspace{-0.3cm}
\end{table}

\section{Key Parameters in Case Studies}
The values of key parameters involved in the case study are listed in Table~\ref{table:Key Parameters}. The basic system parameters include the base power $S_B$ and the short-circuit injection current $\textrm{I}_c$ of IBR. To effectively approximate the SCC constraints, the parameter $\nu$ used for partitioning sample intervals must be properly calibrated, ensuring that the actual SCC levels at all buses are accurately characterized and remain no lower than $\textrm{I}_{b_{\textrm{lim}}}$. For market clearing, the maximum bidding multiplier of strategic players is assumed to be $\overline{\upbeta}^\textrm{m}_{\hat{g}}$ and $\overline{\upbeta}^\textrm{SCC}_{\hat{g},b}$ times the true offers to capture potential market power. Solving the bilevel model requires balancing trade-offs among different objectives, so the weighting parameter W must be carefully selected.

\section*{Acknowledgment}
This work was supported by MICIU/AEI/10.13039/501100011033 and ERDF/EU under grant PID2023-150401OA-C22, as well as by the Madrid Government (Comunidad de Madrid-Spain) under the Multiannual Agreement 2023-2026 with Universidad Politécnica de Madrid, `Line A - Emerging PIs' (grant number: 24-DWGG5L-33-SMHGZ1). The work of Peng Wang was also supported by China Scholarship Council under grant 202408500065.

%\section*{Acknowledgment}
%The authors would like to express sincere gratitude to 

%\IEEEtriggeratref{16}


\begin{thebibliography}{99}

\bibitem{tleis20197}
N.~Tleis, “Short Circuit Analysis Techniques in Large-Scale AC Power Systems,” \textit{Power Systems Modelling and Fault Analysis}, pp.~597–664, 2019.

\bibitem{jia2017review}
J.~Jia, G.~Yang and A.~H.~Nielsen, “A Review on Grid-Connected Converter Control for Short-Circuit Power Provision under Grid Unbalanced Faults,” \textit{IEEE Transactions on Power Delivery}, vol.~33, no.~2, pp.~649–661, 2017.

\bibitem{dozein2018system}
M.~G.~Dozein, P.~Mancarella, T.~K.~Saha and R.~Yan, “System Strength and Weak Grids: Fundamentals, Challenges, and Mitigation Strategies,” \textit{2018 Australasian Universities Power Engineering Conference (AUPEC)}, pp.~1–7, 2018.

\bibitem{huuhtanen2024system}
T.~Huuhtanen, M.~G.~Dozein, S.~P.~Oe and D.~Gheorghe, “System Strength Monitoring: Definition, Metrics and Use Cases Applied to Modern System Operation,” \textit{IET Conference Proceedings (CP876)}, vol.~2024, no.~5, pp.~361–365, 2024.

\bibitem{shao2020power}
W.~Shao, R.~Wu, L.~Ran, H.~Jiang, P.~A.~Mawby, D.~J.~Rogers, T.~C.~Green, T.~Coombs, K.~Yardley, D.~Kastha \textit{et al.}, “A Power Module for Grid Inverter with In-Built Short-Circuit Fault Current Capability,” \textit{IEEE Transactions on Power Electronics}, vol.~35, no.~10, pp.~10567–10579, 2020.

\bibitem{jia2018synchronous}
J.~Jia, G.~Yang, A.~H.~Nielsen, E.~Muljadi, P.~Weinreich-Jensen and V.~Gevorgian, “Synchronous Condenser Allocation for Improving System Short Circuit Ratio,” \textit{2018 5th International Conference on Electric Power and Energy Conversion Systems (EPECS)}, pp.~1–5, 2018.

\bibitem{chu2024pricing}
Z.~Chu, J.~Wu and F.~Teng, “Pricing of Short Circuit Current in High IBR-Penetrated System,” \textit{Electric Power Systems Research}, vol.~235, p.~110690, 2024.

\bibitem{chu2021short}
Z.~Chu and F.~Teng, “Short Circuit Current Constrained UC in High IBG-Penetrated Power Systems,” \textit{IEEE Transactions on Power Systems}, vol.~36, no.~4, pp.~3776–3785, 2021.

\bibitem{matamala2025strategic}
C.~Matamala and G.~Strbac, “Strategic Bidding in the Frequency-Containment Ancillary Services Market,” \textit{Applied Energy}, vol.~401, pp.~126811, 2025.

\bibitem{shukla2011analysis}
U.~K.~Shukla and A.~Thampy, “Analysis of Competition and Market Power in the Wholesale Electricity Market in India,” \textit{Energy Policy}, vol.~39, no.~5, pp.~2699–2710, 2011.

\bibitem{borenstein1995market}
S.~Borenstein, J.~Bushnell, E.~Kahn and S.~Stoft, “Market Power in California Electricity Markets,” \textit{Utilities Policy}, vol.~5, no.~3–4, pp.~219–236, 1995.

\bibitem{wolak2003measuring}
F.~A.~Wolak, “Measuring Unilateral Market Power in Wholesale Electricity Markets: The California Market, 1998–2000,” \textit{American Economic Review}, vol.~93, no.~2, pp.~425–430, 2003.

\bibitem{sweeting2007market}
A.~Sweeting, “Market Power in the England and Wales Wholesale Electricity Market 1995–2000,” \textit{The Economic Journal}, vol.~117, no.~520, pp.~654–685, 2007.

\bibitem{hellmer2009evaluation}
S.~Hellmer and L.~W{\aa}rell, “On the Evaluation of Market Power and Market Dominance—the Nordic Electricity Market,” \textit{Energy Policy}, vol.~37, no.~8, pp.~3235–3241, 2009.

\bibitem{bigerna2016renewable}
S.~Bigerna, C.~A.~Bollino and P.~Polinori, “Renewable Energy and Market Power in the Italian Electricity Market,” \textit{The Energy Journal}, vol.~37, no.~2\_suppl, pp.~123–144, 2016.

\bibitem{lin2022review}
X.~Lin, B.~Wang, Z.~Xiang and Y.~Zheng, “A Review of Market Power-Mitigation Mechanisms in Electricity Markets,” \textit{Energy Conversion and Economics}, vol.~3, no.~5, pp.~304–318, 2022.

\bibitem{li2004market}
T.~Li, M.~Shahidehpour and A.~Keyhani, “Market Power Analysis in Electricity Markets Using Supply Function Equilibrium Model,” \textit{IMA Journal of Management Mathematics}, vol.~15, no.~4, pp.~339–354, 2004.

\bibitem{conejo2020complementarity}
A.~J.~Conejo and C.~Ruiz, “Complementarity, Not Optimization, Is the Language of Markets,” \textit{IEEE Open Access Journal of Power and Energy}, vol.~7, pp.~344–353, 2020.

\bibitem{zeng2014solving}
B.~Zeng and Y.~An, “Solving Bilevel Mixed Integer Program by Reformulations and Decomposition,” \textit{Optimization Online}, pp.~1–34, 2014.

\bibitem{haghighat2018bilevel}
H.~Haghighat and B.~Zeng, “Bilevel Mixed Integer Transmission Expansion Planning,” \textit{IEEE Transactions on Power Systems}, vol.~33, no.~6, pp.~7309–7312, 2018.

\bibitem{jokar2022bilevel}
H.~Jokar, B.~Bahmani-Firouzi and M.~Simab, “Bilevel Model for Security-Constrained and Reliability Transmission and Distribution Substation Energy Management Considering Large-Scale Energy Storage and Demand Side Management,” \textit{Energy Reports}, vol.~8, pp.~2617–2629, 2022.

\bibitem{zhang2017bilevel}
X.~Zhang, D.~Shi, Z.~Wang, Z.~Yu, X.~Wang, D.~Bian and K.~Tomsovic, “Bilevel Optimization Based Transmission Expansion Planning Considering Phase Shifting Transformer,” \textit{2017 North American Power Symposium (NAPS)}, pp.~1–6, 2017.

\bibitem{pan2020bi}
G.~Pan, W.~Gu, H.~Qiu, Y.~Lu, S.~Zhou and Z.~Wu, “Bi-Level Mixed-Integer Planning for Electricity-Hydrogen Integrated Energy System Considering Levelized Cost of Hydrogen,” \textit{Applied Energy}, vol.~270, p.~115176, 2020.

\bibitem{zhu2024optimal}
R.~Zhu, J.~Jiang, J.~Sun, D.~Guo, M.~Wei and Z.~Zhao, “Optimal Planning of Integrated Electricity-Natural Gas Distribution Systems with Hydrogen Enriched Compressed Natural Gas Operation,” \textit{International Journal of Electrical Power \& Energy Systems}, vol.~161, p.~110174, 2024.

\bibitem{huppmann2018exact}
D.~Huppmann and S.~Siddiqui, “An Exact Solution Method for Binary Equilibrium Problems with Compensation and the Power Market Uplift Problem,” \textit{European Journal of Operational Research}, vol.~266, no.~2, pp.~622–638, 2018.

\bibitem{ye2019incorporating}
Y.~Ye, D.~Papadaskalopoulos, J.~Kazempour and G.~Strbac, “Incorporating Non-Convex Operating Characteristics into Bi-Level Optimization Electricity Market Models,” \textit{IEEE Transactions on Power Systems}, vol.~35, no.~1, pp.~163–176, 2019.

\bibitem{Code}
P.~Wang, “Repository of Strategic Bidding in Markets for Short-Circuit Current Ancillary Services,” 2025. [Online]. Available: \url{https://github.com/pwang30/SCC_UC_Bilevel}

\bibitem{najman2021linearization}
J.~Najman, D.~Bongartz and A.~Mitsos, “Linearization of McCormick Relaxations and Hybridization with the Auxiliary Variable Method,” \textit{Journal of Global Optimization}, vol.~80, no.~4, pp.~731–756, 2021.

\bibitem{shahidehpour2004communication}
M.~Shahidehpour and Y.~Wang, “Communication and Control in Electric Power Systems: Applications of Parallel and Distributed Processing”. Hoboken, NJ: John Wiley \& Sons, 2004.

%\bibitem{najman2021linearization}
%J.~Najman, D.~Bongartz and A.~Mitsos, “Linearization of McCormick Relaxations and Hybridization with the Auxiliary Variable Method,” \textit{Journal of Global Optimization}, vol.~80, no.~4, pp.~731–756, 2021.

%\bibitem{cardonha2025recursive}
%C.~Cardonha, A.~Raghunathan, D.~Bergman and C.~Nohra, “Recursive McCormick Linearization of Multilinear Programs,” \textit{INFORMS Journal on Computing}, 2025.

\bibitem{reichert1974computation}
K.~Reichert and N.~Leon, “Computation Methods and Models for Investigating the Stability of Large Synchronous Machines,” \textit{Brown Boveri Review}, vol.~11, pp.~480–487, 1974.


\bibitem{rebours2007survey}
Y.~G.~Rebours, D.~S.~Kirschen, M.~Trotignon, and S.~Rossignol, “A Survey of Frequency and Voltage Control Ancillary Services—Part II: Economic Features,” \textit{IEEE Transactions on Power Systems}, vol.~22, no.~1, pp.~358--366, 2007.

\bibitem{wang2025analyzing}
P.~Wang, Z.~Li, and L.~Badesa, “Analyzing the Impact of Demand Response on Short-Circuit Current via a Unit Commitment Model,” 2025. [Online]. Available: \url{https://arxiv.org/abs/2511.00296}

\end{thebibliography}
\end{document}